\documentclass[journal]{IEEEtran}

\usepackage{cite}
\usepackage{amsmath,amssymb,amsfonts}
\usepackage{algorithm,algorithmic}
\usepackage{graphicx}
\usepackage{textcomp}
\usepackage{xcolor}
\usepackage{hyperref}
\usepackage{siunitx}
\usepackage{diagbox}
\usepackage[caption=false,font=footnotesize]{subfig}
\def\BibTeX{{\rm B\kern-.05em{\sc i\kern-.025em b}\kern-.08em
		T\kern-.1667em\lower.7ex\hbox{E}\kern-.125emX}}

\begin{document}
\addtolength{\textwidth}{0.15in}%

\title{Rechargeable UAV Trajectory Optimization for Real-Time Persistent Data Collection of \\ Large-Scale Sensor Networks}

\author{Rui Wang, \IEEEmembership{Student Member, IEEE}, Deshi Li, Qingqing Wu, \IEEEmembership{Senior Member, IEEE},\\ Kaitao Meng, \IEEEmembership{Member, IEEE}, Boning Feng, and Lele Cong
\thanks{An earlier version of this paper was presented in part at the IEEE International Conference on Communications Workshops, Denver, USA, in June 2024 \cite{wang2024rechargeable}.}
\thanks{R. Wang, D. Li, B. Feng, and L. Cong are with the Electronic Information School, Wuhan University, Wuhan 430072, China (email: ruiwang@whu.edu.cn; dsli@whu.edu.cn; burningf@163.com; conglele@whu.edu.cn).}
\thanks{Q. Wu is with the Department of Electronic Engineering, Shanghai Jiao	Tong University, 200240, China (e-mail: qingqingwu@sjtu.edu.cn).}
\thanks{K. Meng is with the Department of Electronic and Electrical Engineering, University College London, London, UK (email: kaitao.meng@ucl.ac.uk).}
}


\maketitle

\begin{abstract}
Unmanned aerial vehicles (UAVs) have received plenty of attention due to their high flexibility and enhanced communication ability, nonetheless, the limited onboard energy restricts UAVs' application on persistent data collection missions in large areas.
In this paper, we propose a rechargeable UAV-assisted periodic data collection scheme, where a UAV is dispatched to periodically collect data from sensor nodes (SNs) in the mission area and charged by a wireless charging platform.
Specifically, the periodic data collection completion time is minimized by optimizing the UAV trajectory to reach the optimal balance among the collection time, flight time, and recharging time.
The formulated problem is non-convex and difficult to solve directly.
To tackle this problem, we divide the main problem into two sub-problems and address them by leveraging successive convex approximation (SCA), bisection search, and heuristic methods. Then, we propose a periodic trajectory optimization algorithm to iteratively solve the two sub-problems to minimize the completion time. 
Furthermore, to deal with the dynamics of SNs, we propose a low-complexity trajectory adjustment strategy, where the trajectory can be maintained or adjusted locally at the SNs change, which significantly mitigates the computation cost of re-optimization.
The simulation results show the superiority and robustness of the proposed scheme and the completion time is on average 39\% and 33\% lower than the two benchmarks, respectively.
\end{abstract}


\section{Introduction}
\IEEEPARstart{T}{he} intelligent connection of everything is one of the key technologies in the future 6G network. 
To realize the intelligence of wireless networks, efficient data collection from widely deployed smart sensor nodes (SNs) is the basis for the development and application of the intelligent Internet-of-Things (IoT) \cite{Ferrag2023edge}.
However, data communication from IoT devices to ground base stations may be unreliable and sometimes even unavailable due to limited transmission power of SNs, poor channel quality caused by environmental obstacles, or destroyed infrastructures caused by natural disasters \cite{Shen2023fair}. 
In this context, driven by the advantages of high flexibility and enhanced communication ability, unmanned aerial vehicles (UAVs) are emerging facilities for efficient communication and sensing to perform data collection \cite{meng2023throughput}, \cite{meng2023multi}.
{\color{black}For instance, the application of the millimeter-wave (mmWave) frequency band is adopted in UAV-assisted communication systems to support the requirements of high throughput and low latency \cite{Zhu2022Multi-UAV}, \cite{Zhu2020Millimeter}, \cite{Zhu20193-D}.}
UAVs configured with wireless connectivity and sensing platforms can be effectively applied in many different scenarios, such as environmental monitoring, disaster rescuing, and military reconnaissance \cite{Lin2019kalman}, \cite{Sun2019A}, \cite{Qin2021Task}.
The UAV can serve as an aerial mobile base station and provide high-speed data transmission by flying to SNs to establish high-quality links, which can achieve reliable and fast data collection in wireless sensor networks (WSNs) \cite{mozaffari2019tutorial}.
However, with the continuous expansion of the 6G network scale, the massive data generated in WSNs has put forward high persistence and real-time requirements for UAV-enabled data collection.

Existing UAV-assisted data collection works mainly focus on mission performance optimization, such as energy efficiency maximization \cite{Baek2020energy}, Age of Information (AoI) minimization \cite{Jia2019age}, and completion time minimization \cite{liu2022uav}, etc.
For example, the authors in \cite{Jia2019age} studied a UAV path planning problem by jointly considering the access sequence of SNs and data collection mode. Then a solution based on dynamic programming was proposed to determine the optimal access sequence of SNs and minimize the average of AoI. 
In \cite{liu2022uav}, the authors considered the time constraint of the sensory data. They proposed a convex optimization and heuristic-based algorithm to optimize the UAV trajectory and then minimize the completion time.
However, these studies all operate under the assumption that the UAV has enough energy to complete the mission in one flight. 
In practice, the endurance of UAVs is generally limited due to the battery constraint (e.g. 30 minutes for a typical rotary-wing UAV \cite{boukoberine2019critical}), which limits UAVs to only temporary and small-scale missions. Without recharging capabilities, UAVs face challenges in being applied to data collection in increasing large-scale networks \cite{li2018uav}.
Furthermore, sensory data is generated periodically and needs to be collected persistently \cite{Nguyen2018scheduling}, which renders it important to charge UAVs such that enabling UAVs to have a longer endurance and a wider working range, thereby enabling persistent data collection applications in large areas.
Therefore, energy supplement is essential in UAV-enable applications, which results in the completion time not only determined by the flight time and data collection time, but also the UAV recharging time.
In addition, the status of SNs may change as the time span increases during continuous acquisition tasks, for example, new SNs could be deployed to extend the monitoring range \cite{xu2021event}, or the original SNs will fail due to energy exhaustion \cite{javaid2019machine}.
If the trajectory is not adjusted in the face of SNs change, the UAV will either be unable to collect the data of new SNs or increase unnecessary flights to failed SNs, on the other hand, re-planning the global trajectory will bring large computational costs and time consumption. 
Therefore, it is crucial to design an efficient trajectory adjustment method to handle the network dynamics.

To ensure persistent data collection, the near-field inductive wireless charging platform can be deployed for UAV recharging due to the advantages of high efficiency and automation, and in this case, the data collection mission includes several parts: flying to SNs, collecting data, and flying to the charging platform for recharging.
To reduce the mission completion time, several conflicting objectives need to be balanced:
Firstly, to reduce the overall time, the UAV can fly faster and collect more SNs at once, but it will increase the energy consumption and recharging time. Conversely, to reduce energy consumption and save charging time, it will inevitably bring longer flight time, since the UAV will fly at a slower speed in order to save energy.
Secondly, to reduce the data collection time, the UAV will fly closer to SNs so as to enjoy better channel quality. However, this will incur higher completion time, since longer flying distance is required for the UAV to fly closer to each SN.
Thirdly, to reduce the flight time, the SNs subset collected by the UAV during each flight and the data collection sequence need to be carefully planned so as to obtain shorter flight distance.
Therefore, it is highly complex to obtain the optimal trajectory to achieve these balances. Due to the above problems, it is inefficient to directly recalculate the solution of the new problem once the SNs change. However, it is worth noting that, typically, only a small number of SNs undergo change, so the optimized trajectory for original setups holds promise for rapid trajectory adjustments amidst dynamic changes, thus reducing the computational cost of re-optimizing the trajectory.

With the above consideration, we propose a rechargeable UAV-assisted periodic data collection scheme, where a rechargeable UAV is dispatched to periodically collect data from SNs in the mission area and provided with energy supplement by a wireless charging platform.
Different from typical one-flight UAV-enabled systems, our proposed scheme balances UAV flight, data collection, and charging for persistent data collection.
To improve the timeliness of data, a periodic data collection completion time minimization problem is formulated. 
Specifically, the completion time is minimized by optimizing the UAV trajectory, the SNs subset selection during each flight, and the data collection sequence. This optimization ensures that the UAV can be recharged before energy depletion and that each SN can upload a targeted amount of data.
However, solving this optimization problem is highly non-trivial since it is non-convex and involves integer variables closely coupled with the UAV trajectory.
To address this issue, we propose a periodic trajectory optimization algorithm based on convex optimization and bisection method to minimize the completion time efficiently.
Then, aiming at the dynamics of SNs in the persistent tasks, we propose a low-complexity trajectory adjustment strategy based on the historical optimized trajectory, which reduces the computational cost of re-trajectory planning, and improves the adaptability and rapid response ability of dynamic scenarios.
The main contributions are summarized as follows:
\begin{itemize}
    \item We propose a rechargeable UAV-assisted periodic data collection scheme, and formulate a periodic data collection completion time minimization problem by optimizing the UAV trajectory, the SNs subset selection, and the data collection sequence. 
    \item To tackle the formulated problem, we decompose it into the UAV data collection trajectory sub-problem and the SN clustering and visiting order sub-problem. 
    For the UAV data collection trajectory sub-problem, it is converted into a convex problem. Then a successive convex approximation (SCA)-based algorithm is proposed to obtain a Karush-Kuhn-Tucker (KKT) solution.
    \item The SN clustering and visiting order sub-problem is transformed into an asymmetric vehicle routing problem, and the completion time is proved non-decreasing with the charging times.
    Then we propose a bisection and heuristic-based algorithm to solve it. Besides, the lower bound of charging times is derived to reduce the solution space. 
    Finally, the completion time is minimized by alternately solving the two sub-problems.
    \item To deal with the dynamics of SNs, we propose a low-complexity trajectory adjustment strategy that can obtain a suboptimal trajectory while greatly reducing the computational cost.
    The trajectory can be maintained or adjusted locally at the SNs change without re-optimization.
\end{itemize}

The rest of the paper is organized as follows. 
Section \uppercase\expandafter{\romannumeral2} discusses the related works.
In Section \uppercase\expandafter{\romannumeral3}, the system model and problem formulation are presented. 
Section \uppercase\expandafter{\romannumeral4} develops the periodic trajectory optimization algorithm.
In Section \uppercase\expandafter{\romannumeral5} presents the solutions in dynamic scenarios.
Section \uppercase\expandafter{\romannumeral6} provides the simulation results.
Section \uppercase\expandafter{\romannumeral7} concludes this paper.

{\color{black}\emph{Notations:} In this paper, scalars are denoted by italic letters, and vectors are denoted by bold-face lower-case letters.
In addition, $\|\mathbf{x}\|$ represent the Euclidean norm of vector $\mathbf{x}$, $|\mathcal{K}|$ denotes the cardinality of set $\mathcal{K}$, and $\lceil a \rceil $ denotes the minimum integer greater than or equal to $a$.}

\section{Related Works}
To address the energy shortage problem of UAVs for persistent data collection missions, the UAV charging technology has become a critical issue in practical applications. 
Depending on the way of energy replenishment, the UAV battery recharging methods can be divided into two categories: 1) electromagnetic field (EMF)-based and 2) non-electromagnetic field (non-EMF)-based recharging. 
In EMF-based UAV recharging, the energy can be transmitted via a magnetic or electric field \cite{zhang2019wireless}. For example, in \cite{ke2017design}, a wireless charging platform based on magnetic induction is designed, which can provide sufficient energy for the UAV automatically and the charging efficiency can reach more than 90\%.
In non-EMF-based UAV recharging, the UAV can be recharged through non-EMF energy sources like laser beams \cite{zhao2020efficiency} or electromagnetic waves \cite{li2021minimizing}. The authors in \cite{MOHAMMADNIA2021107283} used an 850 nm laser source with an emission power of 600 W for UAV charging.
However, the laser beams and electromagnetic waves charging require a line-of-sight path from the energy source to the UAV, which may be hard to guarantee due to the obstruction of buildings or obstacles in the mission area.  
On the other hand, due to the advantages of high efficiency and stability, the EMF-based UAV recharging technology is more feasible in practical applications.

Combined with UAV energy replenishment, UAV-assisted wireless communication networks have been further studied \cite{li2019rechargeable} \cite{chen2022trajectory}.
In \cite{zhu2021efficient}, the authors proposed a cooperative trajectory planning scheme, where a truck carrying backup batteries moves along with the UAV for data collection. The mission area is first divided into multiple subregions to decide the UAV hovering position, and then the trajectories of the UAV and truck are formulated as a coordinated traveling salesman problem (TSP), which is solved by a three-step trajectory planning algorithm heuristically.
In \cite{zhu2022aerial}, inspired by the aerial refueling scheme, the mission UAVs can be recharged by charging UAVs on the fly. To minimize the mission time, the deep reinforcement learning (DRL) based algorithm is proposed to schedule the flying path and charging process of charging UAVs.
In \cite{zhang2021energy}, the authors considered the UAV-assisted IoT network where a single UAV is powered by solar energy and charging stations for providing communication services. The action-confined on-policy and off-policy reinforcement learning approaches are proposed to optimize the UAV’s trajectory by jointly considering the average data rate, the total energy consumption, and the fairness of coverage for the IoT devices. 
It can be found that most of the above works mainly focus on UAV charging mechanisms in the wireless network.
However, the rechargeable UAV-assisted real-time persistent data collection of large-scale sensor networks is under investigation.
Considering the UAV collecting data within the SN communication coverage during flight, rather than adhering to a binary flying status, introduces various technical complexities in UAV trajectory optimization.
In addition, in the face of SNs change, solving the formulated complex optimization problem repeatedly will bring huge computational cost and time consumption, especially in large-scale networks.

Therefore, an inductive wireless charging platform is deployed for UAV charging due to the advantages of high efficiency and automation, based on this, we propose the rechargeable UAV-assisted periodic data collection scheme and study how to minimize the completion time by optimizing the UAV trajectory and deal with dynamic scenarios.

\section{System Model and Problem Formulation}
\begin{figure}[t]
    \centering{\includegraphics[width=\columnwidth]{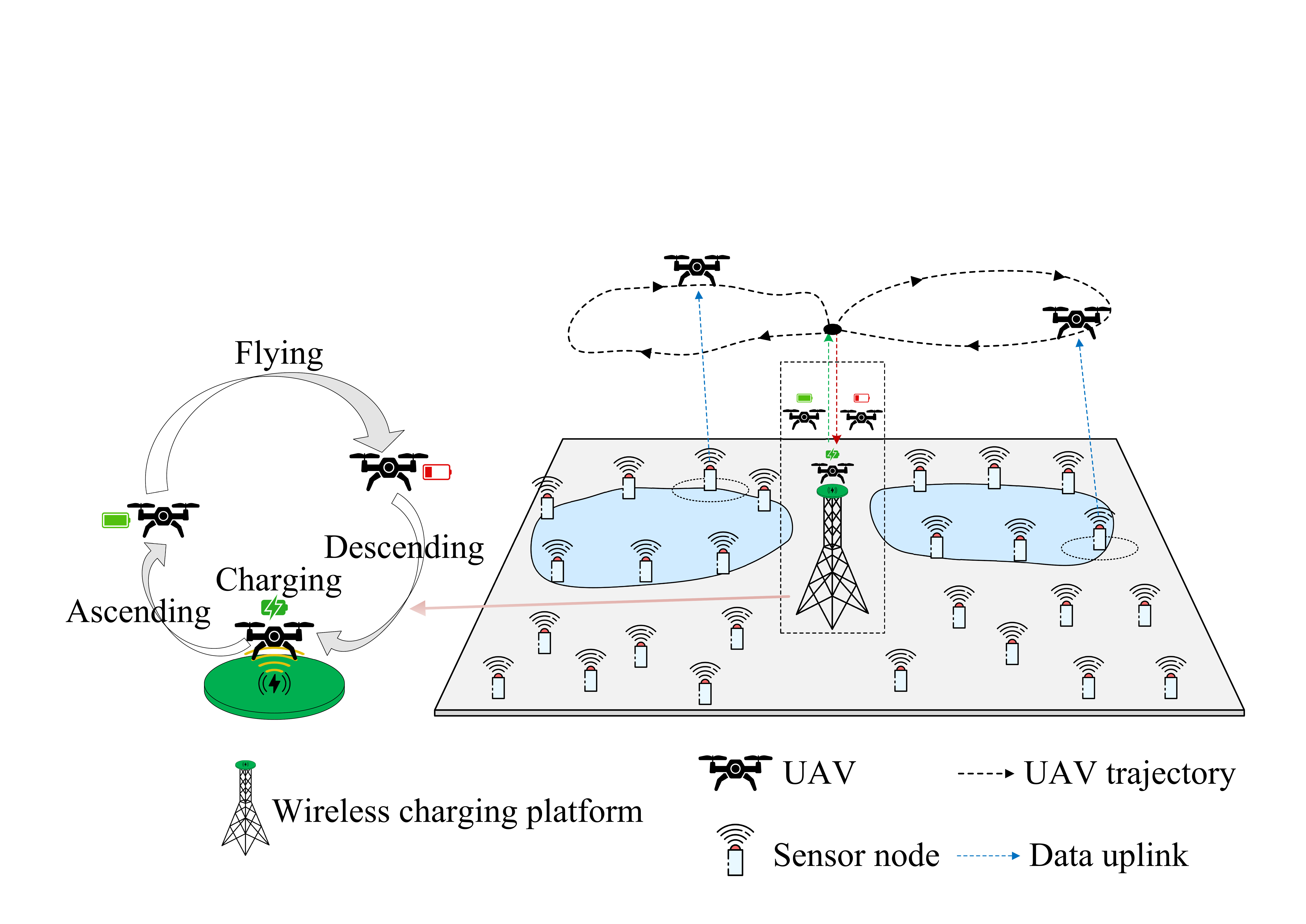}}
    \vspace{-3.5mm}
    \caption{Rechargable UAV-assisted periodic data collection.}
    \label{sys}
\end{figure}

As shown in Fig. \ref{sys}, consider a wireless sensor network with $ K $ SNs, denoted by the set $ \mathcal{K} = \{1,2,...,K\} $, the Cartesian coordinates of which are known and fixed at $ \mathbf{w}_k\in \mathbb{R}^{2\times 1}, k \in \mathcal{K} $. A UAV is dispatched to collect data from all SNs and subsequently returns to the wireless charging platform that is located at $ \mathbf{s}\in \mathbb{R}^{2\times 1} $ for recharging due to limited energy. During one round of data collection from all SNs, suppose that UAV needs to return $N$ times for recharging, {\color{black}thus the SNs can be partitioned into $N$ non-overlapping clusters, i.e., $ G_n \subset \mathcal{K}, 1 \leq n \leq N, \bigcup_{n}G_n=\mathcal{K}, G_{n_1} \cap G_{n_2}=\emptyset,1 \leq n_1 \neq n_2 \leq N $.} Here, $N$ is a variable to be optimized. 
Note that the cluster is a subset of SNs in the logical sense, not the spatial sense.
After collecting the data of all nodes in one round, the next round of data collection is cycled in the same way.

\subsection{System Model}
In this work, the UAV is assumed to fly at a constant altitude $H$ corresponding to authority regulations and safety considerations. In the collection process of a cluster of SNs, the UAV first takes off from the charging platform to the flight altitude, then flies to target SNs to collect data, finally returns to the charging position and lands on the platform for recharging. The UAV is assumed to ascend and descend in constant speed $V_a$, then the time of ascending and descending is
\begin{equation}
    T_{n,ad} = 2\frac{H-H_{C}}{V_a},
\end{equation}
where $H_{C}$ denotes the altitude of charging platform. Therefore, the corresponding energy consumption is $E_{n,ad}=P_a(V_a)T_{n,ad}$. $P_a$ is a function of UAV vertical speed \cite{virgili2022cost},
\begin{equation}
    P_a(V_a) = P_0+\frac{1}{2}WV_a+\frac{1}{2}W\sqrt{V_a^2+\frac{W}{2\rho A}},
\end{equation}
where $W$ is the UAV weight, $\rho$ and $A$ denote the air density and rotor disc area, respectively.

{\color{black}
During the data collection process, the UAV location projected onto the horizontal plane at the time instant $t$ is denoted by $ \mathbf{q}(t)\in \mathbb{R}^{2\times 1} $.
The communication links between the UAV and the SN are assumed to be dominated by the line-of-sight (LoS) component for the elevated UAV \cite{Muruganathan2021}, thus the multipath components are negligible.
Note that in practice, the UAV-ground channel is more likely to have the LoS link as compared to the terrestrial ground-ground channels.
The extension to the non-LoS and multipath fading channels will be left as our future work.
Furthermore, the Doppler effect due to the UAV mobility is assumed to be perfectly compensated. 
Hence, it is assumed that the aerial-ground channel follows the free-space path loss model, and the channel power gain from the SN $k$ to the UAV can be modeled as $h_k(t)=\beta_0 d_k(t)^{-2}= \frac{\beta_0}{H^2+\Vert \mathbf{q}(t)-\mathbf{w}_k \Vert^2}$, where $\beta_0$ is the reference channel power gain at 1m distance.
Therefore, the transmission rate to the UAV for the data collected from the SN $k$ is given by,
\begin{equation}
    R_k(t) = B \log_2\left(1+\frac{P_t \beta_0}{\sigma^2 (H^2+\Vert \mathbf{q}(t)-\mathbf{w}_k \Vert^2)}\right),
\end{equation}
where $B$ is the available bandwidth, $P_t$ is the transmission power of SNs, $\sigma^2$ is the power of channel noise.} To guarantee the successful decoding and quality of service, the signal-to-noise-ratio (SNR) at the UAV, defined by $\frac{P_t \beta_0}{\sigma^2 (H^2+\Vert \mathbf{q}(t)-\mathbf{w}_k \Vert^2)}$, is required to greater than a pre-specified threshold \cite{liao2022energy}. Since SNR is monotonically decreasing with the distance between UAV and SNs, this communication requirement can be satisfied by a distance constraint $\Vert \mathbf{q}(t)-\mathbf{w}_k \Vert \leq d_{th} $, which means that the UAV can only receive data from SN $k$ when it locates in a circular disc with center $\mathbf{w}_k$ and radius $d_{th}$, which is named as communication coverage hereafter.

The UAV trajectory in SNs' communication coverage is represented by $M+1$ waypoints ${\{\mathbf{q}[m]\}}_{m=0}^{M}$ and $M$ time slots ${\{t[m]\}}_{m=1}^{M}$. The length of each segment is sufficiently small, where the distance between the UAV and each SN is approximately unchanged to facilitate the trajectory design.
{\color{black}Thus the flight distance of each line segment is limited by $\Delta_{max}$, which could be an appropriately chosen value such that $\Delta_{max} \ll H$.} 
The waypoints of the SN $l$ of the cluster $G_n$ are denoted by ${\{\mathbf{q}_{n,l}[m]\}}_{m=0}^{M}$, and the corresponding time slots ${\{t_{n,l}[m]\}}_{m=1}^{M}$. 
{\color{black}Define $b_{n,l}[m]$ as the indicator of whether the UAV can receive the data from SN, where $b_{n,l}[m] = 1$ if the UAV's $m$-th waypoint locates in the communication coverage of the SN $l$ of the cluster $G_n$, i.e., $\Vert \mathbf{q}_{n,l}[m]-\mathbf{w}_{n,l} \Vert \leq d_{th}$; otherwise $b_{n,l}[m] = 0$.
Then the UAV trajectory within the SNs' communication coverage should satisfy the following constraint
\begin{equation}
    \begin{aligned}
        &\sum_{m=1}^{M} b_{n,l}[m] t_{n,l}[m] B \log_2 \left(1+\frac{\gamma_0}{H^2+\Vert \mathbf{q}_{n,l}[m]-\mathbf{w}_{n,l} \Vert^2}\right) \\
        &\geq Q_{n,l}, \forall n, \forall l \in G_n,
    \end{aligned}
    \label{rate}
\end{equation}}
where $Q_{n,l}$ is the given SNs' targeting communication requirement and $\gamma_0 \triangleq \frac{P_t \beta_0}{\sigma^2}$. {\color{black}Furthermore, the communication time of data collection of $G_n$ is given by $T_{n,com}=\sum_{l \in G_n}\sum_{m=1}^{M} t_{n,l}[m]$.}

The propulsion power of the UAV is a function of the horizontal speed \cite{zeng2019energy},
\vspace{-1.5mm}
\begin{equation}
    \begin{aligned}
        P(V)= & P_0\left(1+\frac{3V^2}{U_{tip}^2}\right)+P_i\left(\sqrt{1+\frac{V^4}{4v_0^4}}-\frac{V^2}{2v_0^2}\right)^{\frac{1}{2}}\\
        & +\frac{1}{2}d_0\rho sAV^3,
    \end{aligned}
    \vspace{-1.5mm}
\end{equation}
where $P_0$ and $P_i$ represent blade profile power and induced power, respectively. $U_{tip}$ is the tip speed of the rotor blade. $v_0$ denotes the mean rotor induced velocity when hovering. $d_0$ and $s$ are the fuselage drag ratio and rotor solidity, respectively. Also, $\rho$ and $A$ denote the air density and rotor disc area, respectively. Let $z_{n,l}[m] \triangleq \Vert \mathbf{q}_{n,l}[m]-\mathbf{q}_{n,l}[m-1] \Vert$, {\color{black}then the energy consumption of data collection of $G_n$ can be obtained,
\vspace{-1.5mm}
\begin{equation}
    E_{n,com} = \sum_{l \in G_n}\sum_{m=1}^{M} P\left(\frac{z_{n,l}[m]}{t_{n,l}[m]}\right) t_{n,l}[m].
    \vspace{-1.5mm}
\end{equation}}

The UAV is assumed to keep a fixed speed $V_f$ between the SNs' communication coverage, let $[\pi_n(1),...,\pi_n(\vert G_n \vert)]$ be the visiting order of SNs in $G_n$, then the flying time of data collection of $G_n$ can be given by
\vspace{-1.5mm}
\begin{equation}
    T_{n,fly} =\\\frac{1}{V_f}\sum_{l=0}^{\vert G_n \vert} \left\Vert \mathbf{q}_{\pi_n(l)}[M]-\mathbf{q}_{\pi_n(l+1)}[0] \right\Vert,
    \vspace{-1.5mm}
\end{equation}
where $\mathbf{q}_{\pi_n(0)}[0]=\mathbf{q}_{\pi_n(0)}[M]=\mathbf{q}_{\pi_n(\vert G_n \vert+1)}[0]=\mathbf{q}_{\pi_n(\vert G_n \vert+1)}[M]=\mathbf{s}$, $\vert G_n \vert$ is SNs' number in $G_n$, and the corresponding energy consumption is $E_{n,fly}=P(V_f)T_{n,fly}$.

After collecting a cluster of SNs, the UAV will be fully charged for the next cluster data collection, and the wireless charging platform adopts constant power charging mode, therefore the charging time is
\vspace{-1.5mm}
\begin{equation}
    T_{n,chg} = \frac{E_{n,tot}}{P_c},
    \vspace{-1.5mm}
\end{equation}
where $E_{n,tot}=E_{n,com}+E_{n,fly}+E_{n,ad}$, $P_c$ denotes the charging power, then the total data collection time in $G_n$ is $T_{n,tot}=T_{n,com}+T_{n,fly}+T_{n,ad}+T_{n,chg}$. 
The key notations used in this paper are summarized in Table \ref{notation}.

\begin{table}[t]
    \centering
    \caption{Notation and Definition}
    \begin{tabular}{|c|c|} \hline
        Notation & Definition \\ \hline \hline
        $H$ & The altitude of the UAV (m) \\ \hline
        $H_c$ & The altitude of the wireless charging station (m) \\ \hline
        $B$ & Channel bandwidth (Hz) \\ \hline
        $P_t$ & The transmit power of the UAV (Watt) \\ \hline
        $\mathbf{s}$ & The (2D) coordinates of the wireless charging station \\ \hline
        $\mathbf{w}_k$ & The (2D) coordinates of the SN $k$ \\ \hline
        $N$ & The number of SN cluster \\ \hline
        $G_n$ & The $n$-th cluster of SNs \\ \hline
        $d_{th}$ & The SNs' communication coverage radius (m) \\ \hline
        $E_{UAV}$ & The UAV's maximum energy (Joule) \\ \hline
    \end{tabular}
    \label{notation}
\end{table}

\subsection{Problem Formulation}
Based on the previous discussions, we present an optimization problem with the objective to minimize the periodic completion time of data collection of all SNs, i.e.,
\begin{align}
    \text{(P1):} &\min_{ \genfrac{}{}{0pt}{2}{\{\mathbf{q}_{n,l}[m]\}, \{t_{n,l}[m]\},} {N, \{x_{n,k}\}, \{\pi_n(l)\}}} \sum_{n=1}^{N}T_{n,tot}\\
    \mathrm{s.t.\ }
    & \text{(4)},\notag\\
    & \Vert \mathbf{q}_{n,l}[m]-\mathbf{w}_{n,l} \Vert \leq d_{th}, \forall n \in [1,N], \forall l \in G_n, \forall m, \tag{9a}\\
    & \Vert \mathbf{q}_{n,l}[m]-\mathbf{q}_{n,l}[m-1] \Vert \leq \min\{\Delta_{max}, V_{max}t_{n,l}[m] \}, \notag\\
    & \forall n \in [1,N], \forall l \in G_n, \forall m \in [1,M], \tag{9b}\\
    & \left\vert \frac{z_{n,l}[m]}{t_{n,l}[m]} - \frac{z_{n,l}[m+1]}{t_{n,l}[m+1]} \right\vert \leq a_{max}, \notag\\
    & \forall n \in [1,N], \forall l \in G_n, \forall m \in [1,M-1], \tag{9c}\\
    & \left\vert \frac{z_{n,l}[1]}{t_{n,l}[1]} - V_f \right\vert \leq a_{max}, \forall n \in [1,N], \forall l \in G_n, \tag{9d}\\
    & \left\vert \frac{z_{n,l}[M]}{t_{n,l}[M]} - V_f \right\vert \leq a_{max}, \forall n \in [1,N], \forall l \in G_n, \tag{9e}\\
    & E_{n,tot} \leq E_{UAV}, \forall n \in [1,N], \tag{9f}\\
    & [\pi_n(1),\pi_n(2),...,\pi_n(\vert G_n \vert)] \in P_n, \forall n \in [1,N], \tag{9g}\\
    & G_n = \{k|x_{n,k}=1, k\in \mathcal{K}\}, \forall n \in [1,N], \tag{9h}\\
    & x_{n,k} \in \{0,1\}, \forall n \in [1,N], \forall k \in [1,K], \tag{9i}\\   
    & \sum_{n=1}^{N}x_{n,k}=1, \forall k \in [1,K], \tag{9j}
\end{align}
where $x_{n,k}$ are binary variables, and $x_{n,k}=1$ denotes SN $k$ belongs to $G_n$, otherwise $x_{n,k}=0$. Constraint in (4) ensures the SNs' communication throughput requirements. (9a) denotes the UAV can only collect data in SNs' communication coverage area. {\color{black}The maximum UAV speed and segment length constraint is given by (9b), where $\Delta_{max}$ is the maximum length of each line segment.}
{\color{black}The UAV speed change should meet the kinematic constraints in (9c)-(9e) in the acceleration or deceleration process, where $a_{max}$ denotes the threshold of the UAV speed change between two consecutive segments.} (9f) represents the UAV energy constraint. In (9g), $P_n$ denotes all possible permutations of visiting order in $G_n$.

Intuitively, the UAV's flight distance is limited by its battery capacity. If the distance between the SN and the wireless charging platform is large, the constraints in (9f) may not be satisfied and thus (P1) may be infeasible. As a result, we should first analyze the feasibility of the problem (P1).

\textbf{Proposition 1:} (P1) is feasible if and only if $ D \leq (E_{UAV}-E_{ad}-E_{com}^{*}(d_{th},Q_k))\frac{V_{f}}{2P(V_{f})}, \forall k $.

\emph{Proof:} Please refer to Appendix A.
\hfill $\blacksquare$

Note that solving the optimization problem (P1) is highly non-trivial since it is non-convex and involves integer variables. To tackle this challenge, the following result is given.

\textbf{Lemma 1}: The optimal UAV data collection trajectory $\{\mathbf{q}_{n,l}[m]\}$, $\{t_{n,l}[m]\}$ can be obtained if given the optimal SN clusters $N^*$, $\{x^*_{n,k}\}$ and visiting order $\{\pi^*_n(l)\}$ in each cluster.

\emph{Proof:} {\color{black}With given optimal SN clusters and SN visiting order in each cluster $N^*$, $\{x^*_{n,k}\}$, and $\{\pi^*_n(l)\}$, 
the total completion time is decided by the UAV trajectories in SNs' communication coverage.} Then the optimal UAV data collection trajectories $\{\mathbf{q}^*_{n,l}[m]\}$, \{$t^*_{n,l}[m]\}$ can be obtained by $N^*$ independent and homogeneous trajectory optimization problems with energy constraints.
\hfill $\blacksquare$

Inspired by Lemma 1, with given SN clusters and SN visiting order, the UAV data collection trajectory can be optimized to obtain the completion time. 
With the given optimized UAV data collection trajectory, the SN clusters and SN visiting order can be further optimized to obtain a more optimal mission completion time. 
Thus, the problem (P1) will be solved by the optimization of two sub-problems, the UAV data collection trajectory optimization sub-problem and the SN clustering and visiting order optimization sub-problem.

\section{UAV Trajectory Optimization for Data Collection Time Minimization}
In this section, we first decompose the formulated problem into two sub-problems. Then we discuss the solutions to the two sub-problems and propose a periodic trajectory optimization algorithm by alternatively optimizing the two sub-problems efficiently to solve the main problem.

\subsection{UAV Data Collection Trajectory Optimization}
For the given SN clusters and SN visiting order in each cluster, the periodic data collection time minimization problem is equivalent to solving $N$ independent and homogeneous sub-problems (P2-\textit{n}) with \textit{n}=1,2,...,\textit{N} in parallel, where $n$th independent sub-problem optimizes UAV data collection trajectory in $n$th cluster of SNs, i.e.,
\begin{align}
    \text{(P2-\textit{n}):} &\min_{\{\mathbf{q}_{n,l}[m]\}, \{t_{n,l}[m]\}} T_{n,tot}\\
    \mathrm{s.t.\ }
    & \text{(4), (9a)-(9f)}. \notag
\end{align}

To simplify the expression, the subscript $n$ in (P2-\textit{n}) is omitted in the following discussion, and the problem can be rewritten as
\begin{align}
    \text{(P3):} &\min_{\{\mathbf{q}_{l}[m]\}, \{t_{l}[m]\}} T_{tot}\\
    \mathrm{s.t.\ }
    & \sum_{m=1}^{M} t_{l}[m] B \log_2(1+\frac{\gamma_0}{H^2+\Vert \mathbf{q}_{l}[m]-\mathbf{w}_{l} \Vert^2}) \geq Q_{l},\notag\\
    & \forall l \in G_n, \tag{11a}\\
    & \Vert \mathbf{q}_{l}[m]-\mathbf{w}_{l} \Vert \leq d_{th}, \forall l \in G_n, \forall m , \tag{11b}\\
    & \Vert \mathbf{q}_{l}[m]-\mathbf{q}_{l}[m-1] \Vert \leq \min\{\Delta_{max}, V_{max}t_{l}[m] \}, \notag\\
    & \forall l \in G_n, \forall m \in [1,M], \tag{11c}\\
    & \left \vert \frac{z_{l}[m]}{t_{l}[m]} - \frac{z_{l}[m+1]}{t_{l}[m+1]} \right \vert \leq a_{max}, \notag\\
    & \forall l \in G_n, \forall m \in [1,M-1], \tag{11d}\\
    & \left \vert \frac{z_{l}[1]}{t_{l}[1]} - V_f \right \vert \leq a_{max}, \forall l \in G_n, \tag{11e}\\
    & \left \vert \frac{z_{l}[M]}{t_{l}[M]} - V_f \right \vert \leq a_{max}, \forall l \in G_n, \tag{11f}\\
    & E_{tot} \leq E_{UAV}, \tag{11g}
\end{align}
where the objective function is expressed as
\begin{equation}
    \begin{aligned}
        T_{tot} =& \left(\frac{1}{V_f} + \frac{P(V_f)}{P_c V_f}\right) \sum_{l=0}^{\vert G_n \vert} \Vert \mathbf{q}_{\pi(l)}[M]-\mathbf{q}_{\pi(l+1)}[0] \Vert \\
        &+ \frac{P_0}{P_c} \sum_{l \in G_n} \sum_{m=1}^{M} \left(\left(1+\frac{P_c}{P_0}\right)t_l[m] + \frac{3}{U_{tip}^2} \frac{z_l^2[m]}{t_l[m]}\right) \\
        &+ \frac{P_i}{P_c} \sum_{l \in G_n} \sum_{m=1}^{M} \left(\sqrt{t_l^4[m]+\frac{z_l^4[m]}{4v_0^4}}-\frac{z_l^2[m]}{2v_0^2}\right)^\frac{1}{2} \\
        &+ \frac{1}{2P_c} d_0 \rho sA \sum_{l=1}^{\vert G_n \vert} \sum_{m=1}^{M} \frac{z_l^3[m]}{t_l^2[m]} + \left(1+\frac{P_a(V_a)}{P_c}\right)T_{ad}.
    \end{aligned}
\end{equation}

However, the objective function (12) and constraints (11a), (11d), (11g) are non-convex. Therefore, we proposed an efficient algorithm to find the high-quality solution to (P3) based on SCA. 
Firstly, we introduce a set of slack variables to tackle the non-convex terms in (P3).
For the objective function in (12) and constraint (11g), we introduce $\{y_l[m]\}$, such that 
\vspace{-1.5mm}
\begin{equation}
    y_l[m] = \left(\sqrt{t_l^4[m]+\frac{z_l^4[m]}{4v_0^4}}-\frac{z_l^2[m]}{2v_0^2}\right)^\frac{1}{2},
    \vspace{-1.5mm}
\end{equation}
which is equivalent to
\vspace{-1.5mm}
\begin{equation}
    \frac{t_l^4[m]}{y_l^2[m]} = y_l^2[m] + \frac{z_l^2[m]}{v_0^2}.
    \vspace{-1.5mm}
\end{equation}
For constraint (9a), we introduce $\{A_l[m]\},\{d_l[m]\}$, such that 
\vspace{-1.5mm}
\begin{equation}
    A_l^2[m] = t_{l}[m] \log_2\left(1+\frac{\gamma_0}{H^2+d_l^2[m]}\right),
    \vspace{-1.5mm}
\end{equation}
\vspace{-1.5mm}
\begin{equation}
    d_l[m] = \Vert \mathbf{q}_{l}[m]-\mathbf{w}_{l} \Vert.
    \vspace{-1.5mm}
\end{equation}
For constraint (9d), we introduce $\{v_l[m]\}$, such that 
\vspace{-1.5mm}
\begin{equation}
    v_l[m] = \frac{z_{l}[m]}{t_{l}[m]}.
    \vspace{-1.5mm}
\end{equation}

With the above manipulations, (P3) can be rewritten equivalently as 
\vspace{-1.5mm}
\begin{align}
    \text{(P4):} &\min_{\genfrac{}{}{0pt}{2}{\{\mathbf{q}_{l}[m]\}, \{t_{l}[m]\}, \{y_{l}[m]\},}{\{A_{l}[m]\},\{d_{l}[m]\},\{v_{l}[m]\}} } T_{tot}\\
    \mathrm{s.t.\ }
    & \frac{t_l^4[m]}{y_l^2[m]} \leq y_l^2[m] + \frac{z_l^2[m]}{v_0^2}, \forall l \in G_n, \forall m \in [1,M], \tag{18a}\\
    & \sum_{m=1}^{M} A_l^2[m] \geq \frac{Q_l}{B}, \forall l \in G_n, \tag{18b}\\
    & \frac{A_l^2[m]}{t_{l}[m]} \leq \log_2\left(1+\frac{\gamma_0}{H^2+d_l^2[m]}\right), \tag{18c} \\
    & \forall l \in G_n, \forall m \in [1,M], \notag \\
    & \Vert \mathbf{q}_{l}[m]-\mathbf{w}_{l} \Vert \leq d_l[m],\forall l \in G_n, \forall m \in [1,M], \tag{18d}\\
    & \vert v_l[m]-v_l[m+1] \vert \leq a_{max}, \tag{18e} \\
    & \forall l \in G_n, \forall m \in [1,M-1], \notag \\
    & v_l[m] \geq \frac{z_l[m]}{t_l[m]},\forall l \in G_n, \forall m \in [1,M-1], \tag{18f} \\
    & \text{(11b), (11c), (11e), (11f), (11g)}. \notag
    \vspace{-1.5mm}
\end{align}

\textbf{Lemma 2:} The optimal solution to (P3) can be always obtained by solving the problem (P4).

\emph{Proof:} Suppose that the optimal solutions to (P3) and (P4) are $P3^*$ and $P4^*$, respectively. The introduced slack variables $\{\{y_{l}[m]\}, \{A_{l}[m]\},\{d_{l}[m]\},\{v_{l}[m]\} \}$ potentially enlarge the feasible region. Therefore, the optimal solution to (P3) provides an upper bound for the solution to (P4), i.e., $P3^* \geq P4^*$. On the other hand, we can always increase $\{ \{A_{l}[m]\}, \{v_{l}[m]\}\} $ and decrease $\{ \{y_{l}[m]\}, \{d_{l}[m]\}\} $ to satisfy all constraints until the strict equality holds. Consequently, the optimal solution $P4^*$ can always be achieved by $P3^*$, i.e. $P3^* \leq P4^*$. Therefore, The optimal solution to (P3) can be always achieved by addressing the problem (P4). 
\hfill $\blacksquare$

However, the constraints in (18a), (18b), (18c), and (18f) are still complicated and non-convex.
For constraints (18a) and (18b), note that $f(x)=x^2$ is a convex function, there is $x^2 \geq x_0^2+2x_0(x-x_0)$, where $x_0$ is the certain value of the variable $x$. Therefore, given any point $y_l^i[m], z_l^i[m]$ and $A_l^i[m]$, the constraints in (18a) and (18b) can be tightly written as
\vspace{-1.5mm}
\begin{equation}
    \begin{aligned}
		\frac{t_l^4[m]}{y_l^2[m]} \leq & {y_l^i}^2[m]+2y_l^i[m]\left(y_l[m]-y_l^i[m]\right) \\
		&- \frac{1}{v_0^2} \left({z_l^i}^2[m]+2z_l^i[m]\left(z_l[m]-z_l^i[m]\right)\right),
    \end{aligned}
\end{equation}
\vspace{-1.5mm}
\begin{equation}
	\sum_{m=1}^{M} {A_l^i}^2[m] + 2A_l^i[m] \left(A_l[m]-A_l^i[m]\right) \geq \frac{Q_l}{B}.
\end{equation}

Similarly, since  $f(x) = \log_2\left(1+\frac{a}{b+x^2}\right)$ is convex regarding $x$, its global linear lower bound is $\log_2\left(1+\frac{a}{b+x_0^2}\right)-\frac{2a}{ln2} \frac{x_0(x-x_0)}{(x_0^2+b)(x_0^2+b+a)}$ by applying the first-order Taylor expansion of $f(x)$. Therefore, the constraint (18c) has a lower bound
\vspace{-1.5mm}
\begin{equation}
    \begin{aligned}
		\frac{A_l^2[m]}{t_{l}[m]} \leq &\log_2\left(1+\frac{\gamma_0}{H^2+{d_l^i}^2[m]}\right) \\
		&-\frac{2\gamma_0}{ln2}\frac{d_l^i[m]\left(d_l[m]-d_l^i[m]\right)}{\left(H^2+{d_l^i}^2[m]\right)+\left(H^2+{d_l^i}^2[m]+\gamma_0\right)}.
    \end{aligned}
    \vspace{-1.5mm}
\end{equation}

Based on convex function $(x+y)^2$ and its lower bound $ 2(x_0+y_0)(x+y)-(x_0+y_0)^2$, constraint (18f) can be tightly written as
\vspace{-1.5mm}
\begin{equation}
    \begin{aligned}
		v_l[m]t_l[m] & = \frac{1}{2}\left(\left(v_l[m]+t_l[m]\right)^2 - v_l^2[m] - t_l^2[m]\right)\\
		& \geq \left(v_l^i[m]+t_l^i[m]\right)\left(v_l[m]+t_l[m]\right) \\
		& \quad\ -\frac{1}{2}\left({v_l^i}^2[m]+{t_l^i}^2[m]\right)^2 - \frac{1}{2}v_l^2[m] - \frac{1}{2}t_l^2[m]\\
		& \geq z_l[m].
    \end{aligned}
    \vspace{-1.5mm}
\end{equation}

Thus, we can obtain an upper-bounded solution to (P4) by addressing the following convex problem:
\vspace{-1.5mm}
\begin{align}
    \text{(P5):}&\min_{\genfrac{}{}{0pt}{2}{\{\mathbf{q}_{l}[m]\}, \{t_{l}[m]\}, \{y_{l}[m]\},}{\{A_{l}[m]\},\{d_{l}[m]\},\{v_{l}[m]\}} } T_{tot}\\
    \mathrm{s.t.\ }
    & \text{(11b), (11c), (11e), (11f), (11g), (18d), (18e), (19)-(22)}. \notag
    \vspace{-1.5mm}
\end{align}

(P5) can be solved using standard convex optimization techniques like CVX. Finally, by successively updating the local point at each iteration via solving (P5),  the algorithm to solve (P3) is obtained and summarized in \textbf{Algorithm 1}.

\begin{algorithm}[t]
    \footnotesize
    \caption{SCA-Based Algorithm for solving (P3)}
    \begin{algorithmic}[1]
		\STATE Initialize $\{\mathbf{q}_l^i[m]\}, \{t_l^i[m]\}$, and set the iteration round $i=0$.
		\REPEAT 
		\STATE Calculate the current slack variables values $\{y_l^i[m]\},\{A_l^i[m]\},\{d_l^i[m]\},\{v_l^i[m]\}$.
		\STATE Solve the convex problem (P5) and obtain the optimal solution $\{\mathbf{q}_l^*[m]\}, \{t_l^*[m]\}$, and denotes the optimal value as $T_{tot}^i$.
		\STATE Update the local optimization variables as $\{\mathbf{q}_l^{i+1}[m]\}=\{\mathbf{q}_l^*[m]\}, \{t_l^{i+1}[m]\}=\{t_l^*[m]\}$.
		\STATE Update $i=i+1$.
		\UNTIL Convergence is attained or maximal iteration round is achieved.
    \end{algorithmic}
\end{algorithm}

\subsection{SN Clustering and Visiting Order Optimization}
Next, for the given UAV data collection trajectories in SNs' communication coverage, the problem is to optimize the SN clusters and the visiting order in each cluster, i.e.,
\vspace{-1.5mm}
\begin{align}
    \text{(P6):}&\min_{N, \{x_{n,k}\}, \{\pi_n(l)\}} \left(\frac{1}{V_f} + \frac{P(V_f)}{P_c V_f}\right)\sum_{n=1}^{N}D_n + NT_0\\ 
    \mathrm{s.t.\ }
    & \text{(9f)-(9j)}, \notag
    \vspace{-1.5mm}
\end{align}
where $D_n = \sum_{l=0}^{\vert G_n \vert} \Vert \mathbf{q}_{\pi_n(l)}[M]-\mathbf{q}_{\pi_n(l+1)}[0] \Vert$, $T_0$ is constant about UAV takeoff and landing. Since $\{\mathbf{q}_{n,l}[m]\}$ and $ \{t_{n,l}[m]\}$ are fixed, and the constraint (9f) can be expressed as
\vspace{-1.5mm}
\begin{equation}
    D_n \leq \frac{V_f}{P(V_f)}(E_{UAV}-E_{n,com}-E_{n,ad}), \forall n \in [1,N].
    \vspace{-1.5mm}
\end{equation}

\textbf{Theorem 1:} Problem (P6) is NP-hard.

\emph{Proof:} 
{\color{black}
The vehicle routing problem (VRP) is deﬁned as follows: ﬁnd an optimal set of tours with minimum cost to connect the depot to $K$ customers with $N$ vehicles, such that every customer is visited exactly once; every vehicle starts and ends its tour at the depot. It is an NP-hard problem \cite{BRAEKERS2016300}.
If the total travelled distance by each vehicle is required less than or equal to the maximum possible travelled distance, the problem is distance-constrained VRP (DVRP) \cite{6031159}. 
Furthermore, if the distance from node $i$ to node $j$ differs from that of node $j$ to node $i$, this problem is called asymmetric DVRP (ADVRP). These VRP variants are also NP-hard problems \cite{laporte1987branch}. 
In (P6), the set of SNs $\mathcal{K}$ corresponds to the set of customers, the wireless charging platform corresponds to the depot, indexed by $0$, the distance matrix can be expressed as $d_{ij}=\Vert \mathbf{q}_{i}[M]-\mathbf{q}_{j}[0] \Vert $ and $d_{ji}=\Vert \mathbf{q}_{j}[M]-\mathbf{q}_{i}[0] \Vert $, $\forall i,j \in \mathcal{K} \cup \{0\}, i \neq j$, where $\mathbf{q}_{0}[0]=\mathbf{q}_{0}[M]=\mathbf{s}$, and the energy consumption constraint of the UAV corresponds to the distance constraint.
Therefore, (P6) is an equivalent instance of the ADVRP with unknown vehicle numbers, which is a more complex version, as a result, (P6) is also NP-hard. }
\hfill $\blacksquare$

Although (P6) is an NP-hard problem that makes it difficult to find an optimal solution, the solution of (P6) can still be accelerated by searching for the optimal $N$ first.

\textbf{Lemma 3:} The optimal solution to problem (P6) is obtained at the minimum of $N$ in its feasible set.

\emph{Proof:} In the case that the constraint (25) is satisfied, suppose that for $N=N_1$, the optimal objective value of (P6) is $p_{N_1}^*$. If $N=N_2$ and $N_2$ is larger than $N_1$, it is equivalent to splitting some loops out of $N_1$ loops, then the total distance will increase. Furthermore, the second term of the objective function in (P6) increases linearly with $N$. Therefore, $p_{N_1}^* \leq p_{N_2}^*$, where $p_{N_2}^*$ is the optimal objective value of (P6) when $N=N_2$. 
\hfill $\blacksquare$

\textbf{Proposition 2:} {\color{black}Define $E_{tot}^{1-flight}$ as the total energy consumption of the UAV to complete data collection of all SNs in one flight, and $(E_{tot}^{1-flight})^*$ represents the minimum value of $E_{tot}^{1-flight}$.} Then the lower bound of the UAV charging times is
$N^{lb} = \left \lceil \frac{(E_{tot}^{1-flight})^*}{E_{UAV}} \right \rceil.$

\emph{Proof: } Please refer to Appendix B.
\hfill $\blacksquare$

Inspired by Lemma 3, the objective function of (P6) is non-decreasing with $N$, therefore the optimal solution can be found by using bisection search over the cluster number $N$ with constraint (25) holds. 
According to Proposition 2, the lower bound of charging times can be used to improve the computational efficiency of bisection.
With given $N$, to tackle the constraint (25), it is added to the objective function through the penalty function method, therefore we can formulate
\vspace{-1.5mm}
\begin{align}
    \text{(P7):}&\min_{\{x_{n,k}\}, \{\pi_n(l)\}} \left(\frac{1}{V_f} + \frac{P(V_f)}{P_c V_f}\right)\sum_{n=1}^{N} D_n + \notag \\
    & \epsilon \sum_{n=1}^{N} \max\left\{0, D_n -\frac{V_f}{P(V_f)}(E_{UAV}-E_{n,com}-E_{n,ad}) \right\}^2\\
    \mathrm{s.t.\ }
    & \text{(9g)-(9j)}, \notag
    \vspace{-1.5mm}
\end{align}
where $\epsilon$ is the penalty factor, which is set to a sufficiently large positive number, such that the second term of the objective function in (P7) will be very large when constraint (25) is violated.
{\color{black}
Denote the optimal objective value of (P7) as $P_N$ with given cluster number $N$,
When $N=K$, it means that the UAV returns to recharge after collecting one SN until all SNs are collected, and the objective value is denoted as $p_K$. When $N \leq K$ and $p_N > p_K$, it means that the UAV cannot complete the data collection of all SNs in $N$ flights since the violation of the constraint (25) causes the objective value to be too large, otherwise, at least a solution can be obtained to make $p_N \leq p_K$. }

Note that (P7) is an equivalent instance of an asymmetric vehicle routing problem, the high-quality approximate solutions can be efficiently found by existing algorithms like a genetic algorithm (GA). 
Therefore, the problem (P6) can be efficiently solved through bisection search over $N$ and solutions of (P7), the algorithm of solving (P6) is summarized in \textbf{Algorithm 2}.

\begin{algorithm}[t]
    \footnotesize
    \caption{Bisection search and GA-based Algorithm for solving (P6)}
    \begin{algorithmic}[1]
		\STATE Initialize $N^{lb} = \left \lceil \frac{(E_{tot}^{1-flight})^*}{E_{UAV}} \right \rceil, N^{ub}=K$, denote the optimal objective value of (P7) is $p_{N}$ with cluster number is $N$.
		\REPEAT 
		\STATE Update $N=\lceil \frac{N^{lb}+N^{ub}}{2} \rceil$.
		\STATE Solving (P7) using a heuristic algorithm like GA, and obtain the optimal solution $\{x_{n,k}\}, \{\pi_n(l)\} $.
		\IF {$p_N \leq p_K$} 
		\STATE Update $N^{ub}=N$.
		\ELSE
		\STATE Update $N^{lb}=N$.
		\ENDIF
		\UNTIL{$N^{ub}-N^{lb} \leq 1$}.
    \end{algorithmic}
\end{algorithm}

\subsection{Periodic Trajectory Optimization Algorithm}
Based on the discussions in Sections \uppercase\expandafter{\romannumeral3}-A and \uppercase\expandafter{\romannumeral3}-B, the periodic trajectory optimization (PTO) algorithm is further proposed to solve the problem (P1).
Corresponding to the two sub-problems, the proposed algorithm consists of two stages. In the first stage, for the given SN clusters and SN visiting order in each cluster, using \textbf{Algorithm 1} to solve (P2-\textit{n}) with \textit{n}=1,2,...,\textit{N} in parallel and find a set of more optimized data collection trajectories. In the second stage, using \textbf{Algorithm 2} to solve (P6) with data collection trajectory obtained in the first stage and find more optimized SN clusters and SN visiting order. The proposed PTO algorithm alternatively performs the two stages until the objective value of (P1) converges, the details are summarized in \textbf{Algorithm 3}.

\begin{algorithm}[t]
    \footnotesize
    \caption{Periodic Trajectory Optimization Algorithm for solving (P1)}
    \begin{algorithmic}[1]
		\STATE Initialize $N^i=K$, calculate the $\{x_{n,k}^i\}, \{\pi_n^i(l)\} $, set $i=0$.
		\REPEAT 
		\STATE Solving the set of problems (P2-\textit{n}) (\textit{n}=1,2,...,$N^i$) with $N^i, \{x_{n,k}^i\}, \{\pi_n^i(l)\} $ by using Algorithm 1 to obtain the optimized solution $\{\mathbf{q}_{n,l}^i[m]\}, \{t_{n,l}^i[m]\}$.
		\STATE Solving (P6) with $\{\mathbf{q}_{n,l}^i[m]\}, \{t_{n,l}^i[m]\}$ by using Algorithm 2, and obtain the optimal solution $N^{i+1}, \{x_{n,k}^{i+1}\}, \{\pi_n^{i+1}(l)\} $.
		\STATE Update $i=i+1$.
		\UNTIL{The objective value of (P1) converges.}
    \end{algorithmic}
\end{algorithm}

\subsection{Analysis of Bounds}
\emph{The upper bound of} (P3): By introducing a set of slack variables, the problem (P3) can be equivalently transformed to (P4) (c.f. Lemma 2). Through the tighter bounds in (19)-(22), the (P4) is transformed into a convex problem (P5). Note that if the constraints of problem (P5) are satisfied, then those for the original problem (P4) are guaranteed to be satisfied as well, but the reverse is not necessarily true. Thus the solution space of (P5) is a subset of that of (P4), and the optimal value of (P5) provides an upper bound to that of (P4) and (P3). By following the analysis in \cite{marks1978general}, it can be shown that Algorithm 1 is guaranteed to converge to at least a solution that satisfies the KKT conditions of the problem (P4).

\emph{The lower bound of} (P6): From Theorem 1, (P6) is proved to be an instance of an ADVRP. Let $G=(V,E)$, where $ V = \{0,1,...,K\} $ denotes the set of nodes, $0$ is the index of the charging platform, $E$ denotes the set of edges, $E={(i,j)\in V \times V: i \neq j}$. 
The TSP formulation can be obtained by adding $N-1$ copies of the charging platform to $V$. Now, there are $N+K$ nodes in the new graph $G'=(V', E')$, where $V'=V \cup \{K+1,...,K+N-1\}$. Then the distance matrix $D'=[d'_{ij}]$ can be obtained as:
\vspace{-1.5mm}
\begin{equation}
    d'_{ij} =
    \begin{cases}
        d_{ij}, i,j \in [0,K], i \neq j \\
        d_{i0}, i \in [1,K], j \in [K+1, K+N-1] \\
        d_{0j}, i \in [K+1, K+N-1], j \in [1,K] \\
        \infty, otherwise
    \end{cases}.
    \vspace{-1.5mm}
\end{equation}
When the distance constraint in (25) is relaxed, the ADVRP can be transformed into an asymmetric TSP as follows \cite{almoustafa2013new}:
\vspace{-1.5mm}
\begin{align}
    &\min_{X} \sum_{(i,j) \in E'} d'_{ij} x_{ij} \\
    \mathrm{s.t.\ }
    & \sum_{i \in V'} x_{ij} = 1, \forall j, \tag{28a} \\
    & \sum_{j \in V'} x_{ij} = 1, \forall i, \tag{28b} \\
    & \sum_{i,j \in U} x_{ij} \leq \vert U \vert -1, \forall U \subseteq V'\setminus {0}, \vert U \vert \geq 2, \tag{28c} 
    \vspace{-1.5mm}
\end{align}
where $x_{ij}=1$ if the edge $(i,j)$ belongs to any tour and $i \neq j$, otherwise it is 0. The constraint (28c) eliminates subtours, where $U$ is any subset of $V'\setminus{0}$, and $|U|$ is the cardinality of the set $U$. Similar to \cite{meng2021space}, the above problem can be converted into a symmetric TSP as follows: Construct the new distance matrix $\widetilde{D} = \begin{bmatrix}
    H & {D'}^T \\
    D' & H
\end{bmatrix} $ , 
where $H$ is a (N+K)-dimensional square matrix and $h_{ij}=\infty, \forall i,j$.
According to \cite{jonker1983transforming}, the optimal solution of the new symmetric problem is equal to that of the original asymmetric problem. When $N=N^{lb}$, the optimal solution of the new symmetric problem provides a lower bound of (P6), which can be obtained by finding the looped minimum 1-tree, if the distance constraint is satisfied, it is the optimal solution of (P6).

{\color{black}
\emph{The lower bound of} (P1): With the given lower bound of charging times $N^{lb}$, the lower bound of (P6) can be obtained by finding the looped minimum 1-tree generated from graph $V'$ by the Lin-Kernighan-Helsgaun (LKH) algorithm \cite{LKH}. With the obtained SN clusters and SN visiting order in each cluster, using Algorithm 1 to optimize the data collection trajectories. Then, the lower bound of (P1) can be obtained.}

\subsection{Complexity Analysis}
{\color{black}
The computational complexity of the proposed PTO algorithm mainly depends on the complexity of Algorithms 1 and 2. 
For Algorithm 1, the complexity by using the interior point method to solve the optimization problem is $\mathcal{O}((L(7M+2))^{3.5})$, where $L(7M+2)$ is the number of variables in (P5). 
Thus it is estimated that the complexity of Algorithm 1 is approximately $\mathcal{O}(I_1(LM)^{3.5})$, where $I_1$ is the number of iterations of Algorithm 1. Then, the complexity of UAV data collection trajectory optimization from line 3 in Algorithm 3 is $\mathcal{O}(I_1(KM)^{3.5})$.
For Algorithm 2, its computational complexity mainly depends on the bisection method and GA. 
The complexity of the bisection is $\mathcal{O}(\log K)$.
The complexity of GA is $\mathcal{O}(I_2PK)$, where $I_2$ is the number of iterations of GA, $P$ is the population size, and $K$ is the chromosome size which corresponds to the number of SNs. 
As a result, the overall computational complexity of the proposed PTO algorithm is approximately $\mathcal{O}(I_1(KM)^{3.5}+I_2PK\log K)$. Since $I_2PK\log K \ll I_1(KM)^{3.5}$, thus the computational complexity can be further approximately $\mathcal{O}(I_1(KM)^{3.5})$.}

\begin{figure}[t]
    \centering
    \subfloat[Disjoint with the optimized trajectory.]{\centering \includegraphics[width=1.8in]{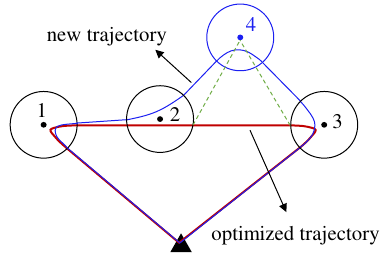}}
    \subfloat[Intersecting the optimized trajectory.]{\centering \includegraphics[width=1.8in]{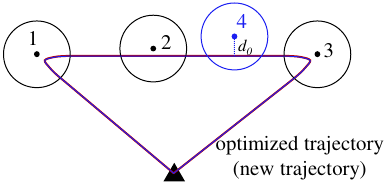}}
    \caption{The location of the new node.}
    \label{node addition}
\end{figure}

\begin{figure}[t]
    \centering
    \subfloat[At the turn of the optimized trajectory.]{\centering \includegraphics[width=1.8in]{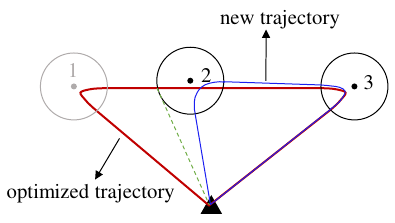}}
    \subfloat[At the straight line of the optimized trajectory.]{\centering \includegraphics[width=1.8in]{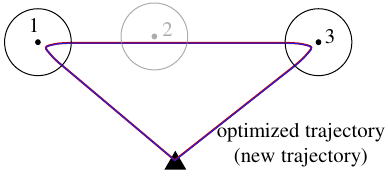}}
    \caption{The location of the failure node.}
    \label{node failure}
\end{figure}

\section{Solutions With Dynamic Scenarios}
In this section, we consider the local optimized trajectory re-planning for dynamic scenarios.  
To reduce the computational cost caused by re-planning the global trajectory, we analyze the local trajectory adjustment strategy under different SNs change scenarios.
Generally, the changes of SNs can be divided into two categories: the SN addition and the SN failure.
In practice, this could correspond to the scenarios where the new SNs are deployed for more extensive and detailed monitoring or the original SNs fail due to fault or power supply. 

\subsection{SN Addition}
\emph{1) Scenario (a)}, The new SN's communication coverage does not intersect the optimized trajectory: 
As illustrated in Fig. \ref{node addition}(a), in this case, the $s_4$ is the new node, then the nearest node to the new node should be found, and insert $s_4$ to the trajectory that minimizes the total length, i.e. the SN visiting order should change from $\{s_1, s_2, s_3\}$ to $\{s_1, s_2, s_4, s_3\}$. 

\textbf{Remark 1:} Let $s_i$ and $s_{i+1}$ be two adjacent SNs on the original trajectory and $s_j$ be a new SN, then there must exist $i$ such that $s_is_j+s_js_{i+1}-s_is_{i+1}$ is minimal. Therefore, inserting $s_j$ between $s_i$ and $s_{i+1}$ can minimize the length of the new trajectory.

Thus, a suboptimal trajectory can be constructed by disconnecting $\overline{s_{2,o}s_{3,i}}$ and then connecting $\overline{s_{2,o}\mathbf{w}_{4}}$ and $\overline{\mathbf{w}_{4}s_{3,i}}$, where $s_{2,o}$ is the exit point of $s_2$'s coverage, and $s_{3,i}$ is the entry point of $s_3$'s coverage on the original optimized trajectory.

However, since the data collection of new nodes will bring additional flight distance and thus increase energy consumption, to make the data collection in the new visiting order still feasible (i.e. to meet the energy consumption constraint of UAV), a tight constraint can be set: 
\begin{equation}
    \begin{aligned}
        &\frac{\Vert s_{2,o}-\mathbf{w}_{4} \Vert + \Vert \mathbf{w}_{4}-s_{3,i} \Vert - \Vert s_{2,o}-s_{3,i} \Vert}{V_f} P(V_f) + \frac{Q}{R_h}P_h\\
        &\leq E_{UAV}-E_{tot},
    \end{aligned}
\end{equation}
where $R_h$ and $P_h$ denote the communication rate and UAV's power when hovering above the SN, respectively, $E_{tot}$ is the total energy consumption of the original optimized trajectory. 
Then, the new trajectory can be optimized with the new visiting order by setting the initial point and final point to $s_{1,i}$ and $s_{3,o}$.

\textbf{Theorem 2:} Given the objective value $T$ of the original problem and the distance $L$ from the new SN to the optimized trajectory, The gap between the total completion time of the constructed suboptimal trajectory and the new optimal trajectory is bounded by $\left(\frac{2L}{V_f}+\frac{Q}{R_h}\right)/T$.

\emph{Proof:} The total completion time of the constructed suboptimal trajectory can be expressed as $T+\Delta T$, $\Delta T \leq \frac{\sqrt{d^2+4L^2}-d}{V_f} + \frac{Q}{R_h}$, where $d=\Vert s_{2,o}-s_{3,i} \Vert$. Assuming that $T'$ is the optimal objective value for the problem with new SN, then the following inequation can be obtained:
\vspace{-1.5mm}
\begin{equation}
    \begin{aligned}
        \frac{T+\Delta T}{T'} \leq \frac{T+\frac{\sqrt{d^2+4L^2}-d}{V_f} + \frac{Q}{R_h}}{T'} < \frac{T+\frac{2L}{V_f} + \frac{Q}{R_h}}{T}.
    \end{aligned}
    \vspace{-1.5mm}
\end{equation}
\hfill $\blacksquare$

Theorem 2 states that if the UAV has enough residual energy to support the data collection of a new SN with the trajectory represented by the green dashed line in Fig. \ref{node addition}(a), the gap between the completion time of the trajectory and the corresponding optimal trajectory is mainly determined by the distance between the new SN and the original optimal trajectory $L$. If the length of the original optimal trajectory is much larger than $L$ or $L$ is small, the gap can be ignored and the constructed trajectory has high quality and low complexity.

\emph{2) Scenario (b)}, The new SN's communication coverage intersects the optimized trajectory:
As illustrated in Fig. \ref{node addition}(b), if the new SN's communication coverage is passed by the optimized trajectory, the UAV's data collection time can be obtained as $T=\frac{2\sqrt{d_{th}^2-d_0^2}}{V_f}$, where $d_0$ is the distance from the new SN to the optimized trajectory. Thus, the amount of data collected by the UAV can be expressed as
\vspace{-1.5mm}
\begin{equation}
    \begin{aligned}
        \tilde{Q} = & B \int_{0}^{T} \log_2\left(1+\frac{\gamma_0}{H^2+(d_{th}-V_ft)^2}\right) dt\\
        = &\frac{2B}{V_f\ln{2}}
        \left(\sqrt{d_{th}^2-d_{0}^2} \ln{\frac{H^2+d_{th}^2+\gamma_0}{H^2+d_{th}^2}} \right.\\
        & +2\sqrt{H^2+d_0^2+\gamma_0}\arctan\frac{\sqrt{d_{th}^2-d_{0}^2}}{2\sqrt{H^2+d_0^2+\gamma_0}}\\
        & \left. -2\sqrt{H^2+d_0^2}\arctan\frac{\sqrt{d_{th}^2-d_{0}^2}}{2\sqrt{H^2+d_0^2}}\right).
    \end{aligned}
    \label{length}
    \vspace{-1.5mm}
\end{equation}

\textbf{Theorem 3:} Given the optimal solution $\Phi^*=\{N^*, \{x^*_{n,k}\}, \{\pi^*_n(l)\}, \{\mathbf{q}^*_{n,l}[m]\}, \{t^*_{n,l}[m]\}\}$ of the original problem, if the new SN satisfies the condition $\tilde{Q} \geq Q$, where $Q$ is the SN's communication throughput requirement, then $\Phi^*$ is also the optimal solution to the new problem.

\emph{Proof:} Since the data collection of the new node means increasing the flight distance and energy consumption of UAV, thus $T' \geq T$, where $T'$ and $T$ denote the total completion time of the new problem and the original problem, respectively. If $\tilde{Q} \geq Q$ is satisfied, then $\Phi^*$ satisfies all the constraints of the new problem and the corresponding objective value is $T$, therefore, $\Phi^*$ is the optimal solution of the new problem.
\hfill $\blacksquare$

Theorem 3 states that if the distance between the new SN and the original optimal trajectory is less than a threshold, which can be solved by equation (\ref{length}) and related to SN throughput requirement and communication coverage, then the optimal solution to the new problem remains consistent with the original solution.

\subsection{SN Failure}
\emph{1) Scenario (a)}, The failure node is at the turn of the optimized trajectory:
As illustrated in Fig. \ref{node failure}(a), the failure node is $s_1$, then it will be deleted directly in this collection, and the new trajectory will link the two adjacent nodes of the failure node. Thus the visiting order changes from $\{s_1, s_2, s_3\}$ to $\{s_2, s_3\}$. 
Since the flight path is shortened by not collecting data from the failed node, the new trajectory is feasible and can be optimized with the new visiting order by setting the initial point and final point to $s$ and $s_{3,o}$.

\emph{2) Scenario (b)}, The failure node is at the straight line of the optimized trajectory: 
As illustrated in Fig. \ref{node failure}(b), the failure node is $s_2$, in this case, similar to the \emph{Scenario (b)} in Section \uppercase\expandafter{\romannumeral5}-A, the original optimized trajectory is still the new optimized trajectory. 
The main reason is that the extra energy is not enough to collect new SN data on the current path.

\section{Simulation and Results Analysis}
\begin{table}[t]
    \centering
    \caption{Parameter Settings}
    \begin{tabular}{|c|c|} \hline
        Parameters & Value \\ \hline \hline
        The wireless charging platform position $\mathbf{s}$ & (2500, 2500) \\ \hline
        The UAV's maximum speed $V_{max}$ & 30 m/s \\ \hline
        The UAV's vertical speed $V_{a}$ & 6 m/s \\ \hline
        The UAV's flight speed $V_{f}$ & 18 m/s \\ \hline
        The UAV's maximum acceleration speed $a_{max}$ & 5 m/s \\ \hline
        The SNs' communication coverage radius $d_{th}$ & [50, 500] m \cite{li2023completion} \\ \hline
        The SNs' communication throughput $Q$ & [1, 250] Mbits \\ \hline
        The UAV's maximum energy $E_{UAV}$ & [50, 175] KJ \\ \hline
        The wireless charging power $P_{c}$ & 150 W \\ \hline
        \color{black}The penalty factor $\epsilon$ in (P7) & \color{black} \num{1.0e11} \\ \hline
    \end{tabular}
    \label{Parameters}
\end{table}

\begin{figure}[t]
    \centering{\includegraphics[width=2.3in]{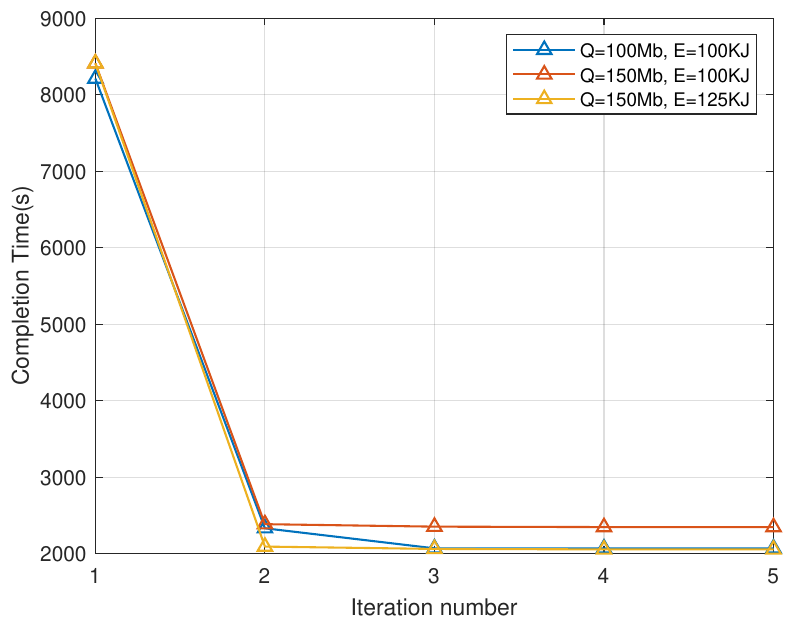}}
    \caption{Convergence of Algorithm 3.}
    \vspace{-3.5mm}
    \label{convergence}
\end{figure}

\begin{figure*}[t]
    \centering
    \subfloat[Greedy-based algorithm (3478.8s).]{\centering \includegraphics[width=2.3in]{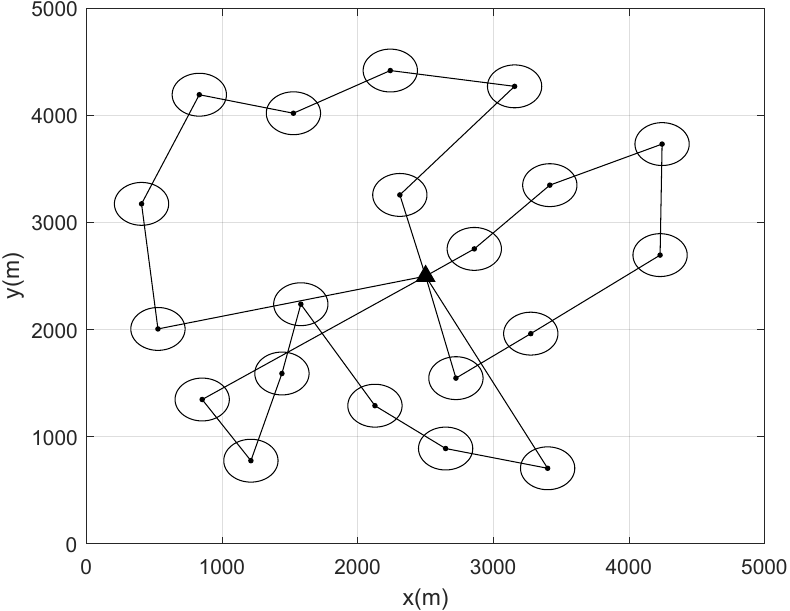}}
    \subfloat[Optimized Hmode algorithm (3214.0s).]{\centering \includegraphics[width=2.3in]{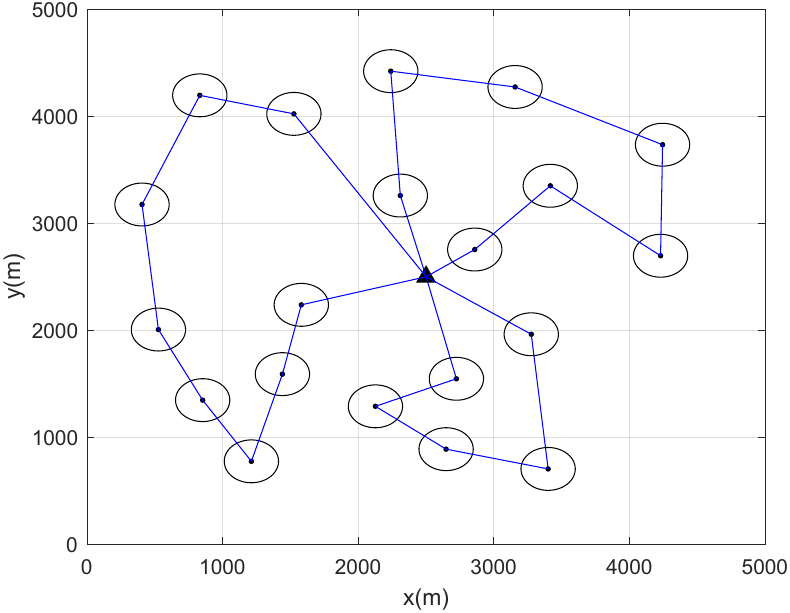}}
    \subfloat[PTO algorithm (2068.6s).]{\centering \includegraphics[width=2.3in]{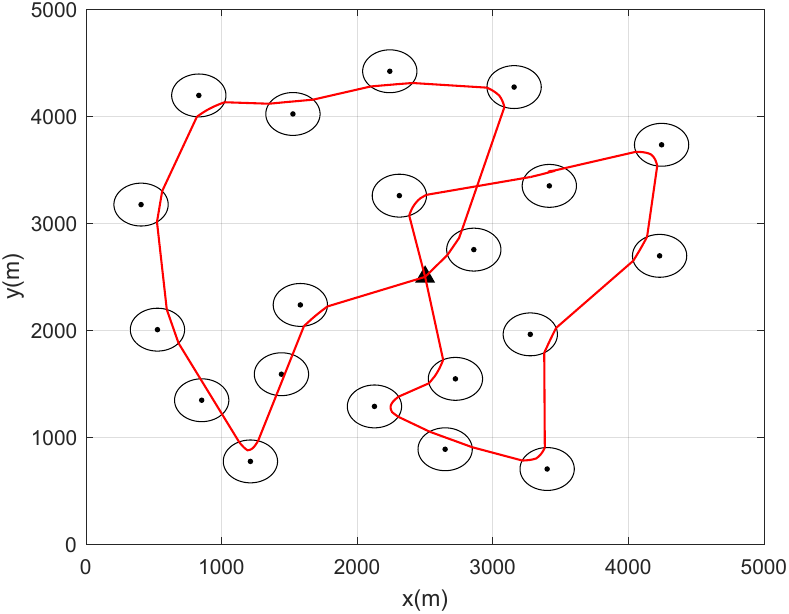}}
    \caption{UAV trajectories with different algorithms ($Q$=100Mbits, $E_{UAV}$=100KJ, $d_{th}=200m$).}
    \label{different trajectories}
\end{figure*}

\begin{figure}[t]
    \centering{\includegraphics[width=2.3in]{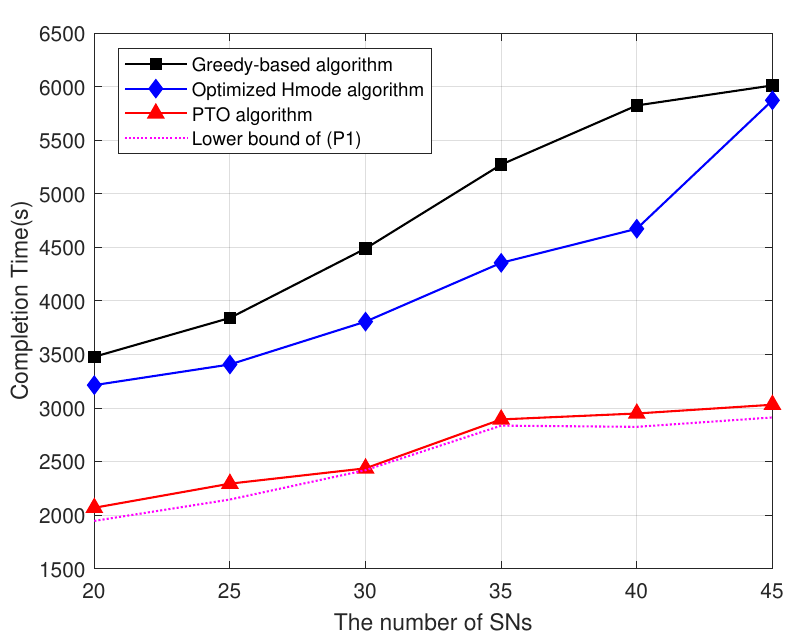}}
    \vspace{-3.5mm}
    \caption{Completion time versus the number of SNs ($Q$=100Mbits, $E_{UAV}$=100KJ, $d_{th}=200m$).}
    \label{TvsSN}
\end{figure}

This section provides numerical results to validate the proposed algorithms. 
The UAV altitude is set as $H$ = 100 m and the charging platform altitude is set as $H_{C}$ = 15 m. 
{\color{black}We consider a sensor network with $K$ = 20 SNs, and the SNs are randomly located in a square area, whose side length is equal to 5 km.
The results except Fig. \ref{TvsSN} obtained in the following are based on one realization of SNs' locations as shown in Fig. \ref{different trajectories}. }  
The total communication bandwidth is $B$ = 1 MHz, the SN's transmit power is set as $P_t$ = 0.1 W, the noise power is -110 dBm, the channel power gain $\beta_0$ at the reference distance of 1 m is set as -60 dB. The UAV energy related parameters are set based on \cite{zeng2019energy}, where $U_{tip}^2=120$, $v_0=4.03$, $d_0=0.6$, $\rho=1.225$, $s=0.05$, $A=0.503$, $P_0=79.86$, $P_i=88.63$ and $W=20$. The other parameters are shown in Table \ref{Parameters}, and the simulation tool is MATLAB.

For evaluation, the proposed algorithm is compared with two benchmarks: a greedy-based algorithm and an optimized Hmode algorithm. In both benchmarks, the UAV collects data by hovering above the SNs. 
The greedy-based algorithm first designs the shortest trajectory so that the UAV can collect all the SNs' data in one flight, and then the UAV returns to recharge when its energy is depleted while flying along the shortest trajectory. Compared to the greedy-based algorithm, the optimized Hmode algorithm has optimal SN clusters and SN visiting orders in each flight.
{\color{black} The details of the benchmarks are summarized in \textbf{Algorithm 4} and \textbf{Algorithm 5}.}

\begin{algorithm}[t]
    \footnotesize
    \color{black}
    \caption{Greedy-based Algorithm}
    \begin{algorithmic}[1]
        \STATE Initialize the SN visiting order $\Gamma$ to the TSP path containing all SNs, the UAV's current point is the charging platform, and the UAV residual energy $E_{res} = E_{UAV}$.
        \REPEAT 
        \STATE Obtain the next uncollected SN $k$ from the visiting order.
        \STATE Calculate the energy consumption for collecting SN $k$: $E_{k}$, and the return energy required from SN $k$: $E_{k,ret}$.
        \IF{$E_{res} \geq E_k + E_{k,ret}$}
        \STATE The UAV flies to collect data of SN $k$, $E_{res}=E_{res}-E_{k}$.
        \ELSE
        \STATE The UAV returns to the charging platform, $E_{res}=E_{UAV}$.
        \ENDIF
        \UNTIL All SNs are collected.
    \end{algorithmic}
\end{algorithm}

\begin{algorithm}[t]
    \footnotesize
    \color{black}
    \caption{Optimized Hmode Algorithm}
    \begin{algorithmic}[1]
        \STATE Initialize the UAV data collection locations for SNs: $\mathbf{q}_{k} = \mathbf{w}_k, k \in \mathcal{K}$.
        \STATE Calculate the data collection time of SNs: $t_k = \frac{Q_k}{R}, k \in \mathcal{K}$, where $B\log_2(1+\gamma_0/H^2)$.   
        \STATE Execute the lines 2 to 10 in Algorithm 2.
        \RETURN Optimized SN clusters and visiting order.
    \end{algorithmic}
\end{algorithm}

First, we investigate the convergence of the proposed PTO algorithm. 
{\color{black}Fig. \ref{convergence} shows the UAV's completion time with the number of iterations in the case of different SN communication requirements ($Q$=100Mbits, 150Mbits) and UAV energy ($E_{UAV}$=100KJ, 125KJ). }
{\color{black}It can be seen that the completion time converges within a few iterations, which demonstrates the efficiency of the proposed algorithm.
The main reason is that the initialization step of algorithm 2 reduces the search space of the cluster number so that the algorithm can quickly iterate to the optimal cluster number.}

\begin{figure}[t]
    \centering{\includegraphics[width=2.3in]{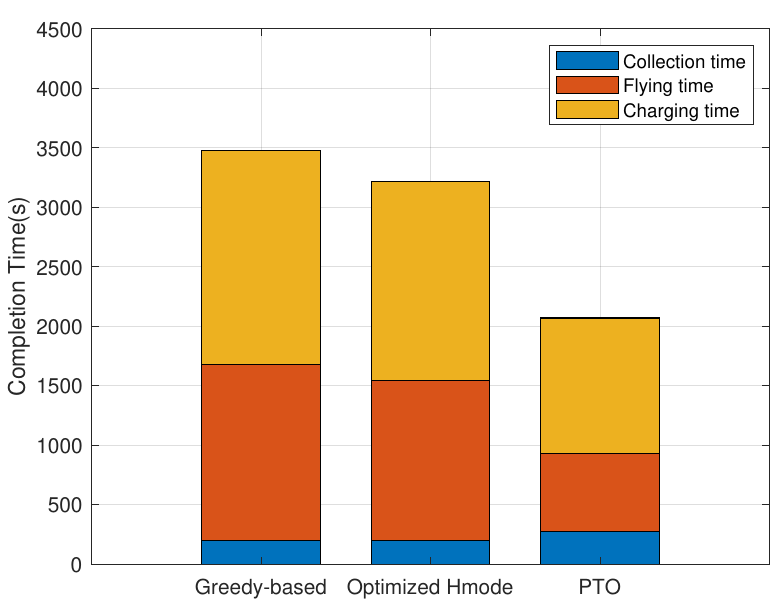}}
    \vspace{-3.5mm}
    \caption{Time allocation with different algorithms ($Q$=100Mbits, $E_{UAV}$=100KJ, $d_{th}=200m$).}
    \label{time alloction}
\end{figure}

\begin{figure}[t]
    \centering{\includegraphics[width=2.3in]{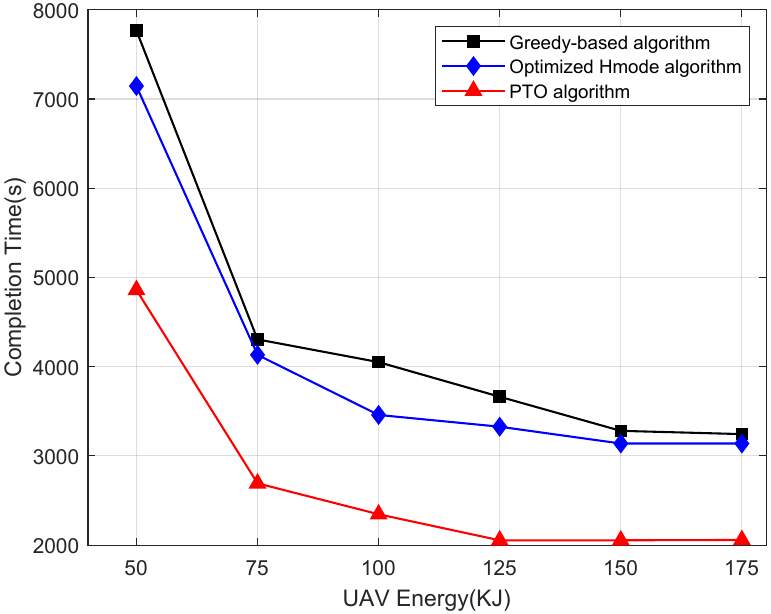}}
    \vspace{-3.5mm}
    \caption{Completion time versus the UAV energy ($Q$=150Mbits, $d_{th}=200m$).}
    \label{TvsE}
\end{figure}

{\color{black}In the following, the results in Fig. \ref{different trajectories}-\ref{TvsSN} are obtained with $Q$=100Mbits, $E_{UAV}$=100KJ, and $d_{th}$=200m.}
Fig. \ref{different trajectories} shows the UAV trajectories obtained by different algorithms, where the black dots and circles are SNs and their communication coverage, and the triangle represents the wireless charging platform. 
Specifically, the completion time of the benchmarks is 68.2\% and 55.4\% higher than that of the proposed algorithm, respectively. 
It can be seen that the trajectory obtained by the proposed algorithm is smoother in SNs' communication coverage and returns less for recharging as compared to that obtained by the benchmarks. 
The main reason is that, in the proposed algorithm, the trajectories in SNs' communication coverage achieved the balance between collection and flight time, based on which, the SN clusters and flight path in each cluster are optimized for shorter flight distance and less energy consumption to obtain the UAV trajectory.
Hence, the complete time achieved by the proposed algorithm is shorter.

Fig. \ref{time alloction} shows the time allocation of data collection, flight, and charging in the total task completion time of the UAV under different algorithms. 
Since the benchmark algorithms both collect data by hovering above the node, they have the same collection time. 
Compared with the greedy-based algorithm, the optimized Hmode algorithm optimizes the SN visiting order, so the flight time is shorter, the flight energy consumption is saved, and the charging time is reduced, resulting in a shorter total completion time. 
Due to the proposed PTO algorithm adopting the method of data collection during the flight within the communication coverage, although the collection time is increased, the collection energy consumption is reduced. 
More nodes are collected at one time, and the total flight distance and flight time are reduced, thus the energy consumption and charging time are reduced, and the completion time is reduced.

\begin{figure}[t]
    \centering{\includegraphics[width=2.3in]{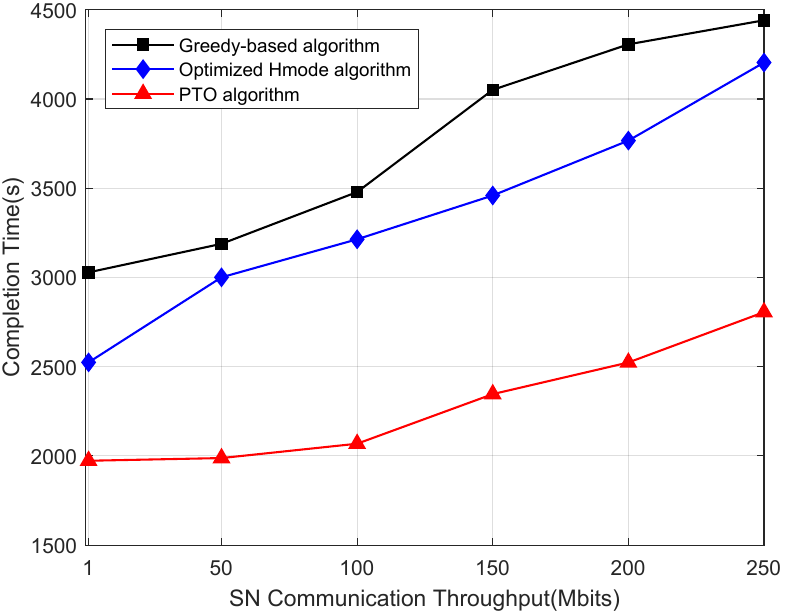}}
    \vspace{-3.5mm}
    \caption{Completion time versus the SN communication throughput requirement ($E_{UAV}$=100KJ, $d_{th}=200m$).}
    \label{TvsQ}
\end{figure}

\begin{table}[t]
    \centering
    \color{black}
    \caption{Execution Time (in seconds) for Different Algorithms}
    \begin{tabular}{|c|c|c|c|c|c|c|} \hline
        Number of SNs & 20 & 25 & 30 & 35 & 40 & 45 \\ \hline 
        Greedy-based & 4.7  & 5.2  & 6.7  & 7.0  & 8.4 & 9.3  \\ \hline
        Optimized Hmode & 17.9  & 18.3  & 19.2  & 19.6  & 20.1 & 23.4  \\ \hline
        PTO & 23.2  & 25.6  & 27.8  & 29.4  & 32.6 & 35.1 \\ \hline
    \end{tabular}
    \label{time}
\end{table}

\begin{figure}[t]
    \centering{\includegraphics[width=2.3in]{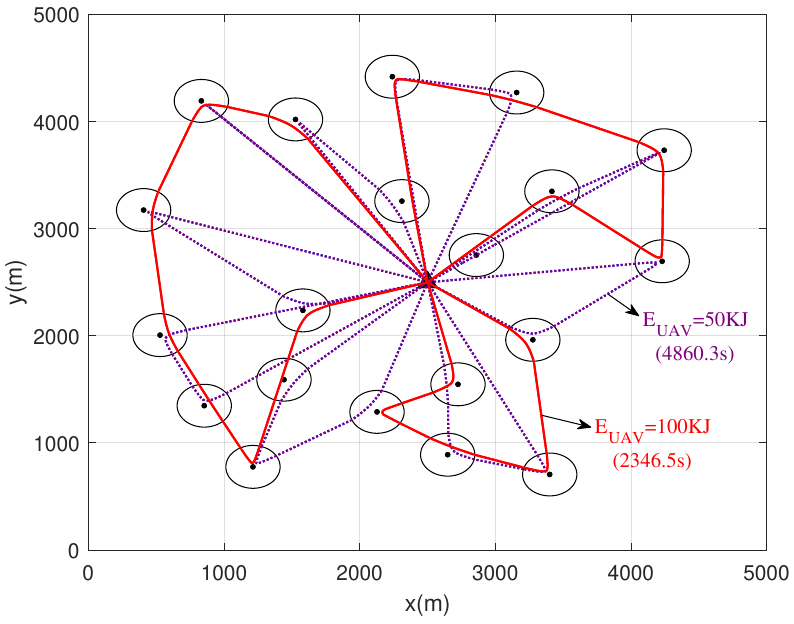}}
    \vspace{-3.5mm}
    \caption{UAV trajectory versus the UAV energy ($Q$=150Mbits, $d_{th}=200m$).}
    \label{trajectory vs E}
\end{figure}

\begin{figure}[t]
    \centering{\includegraphics[width=2.3in]{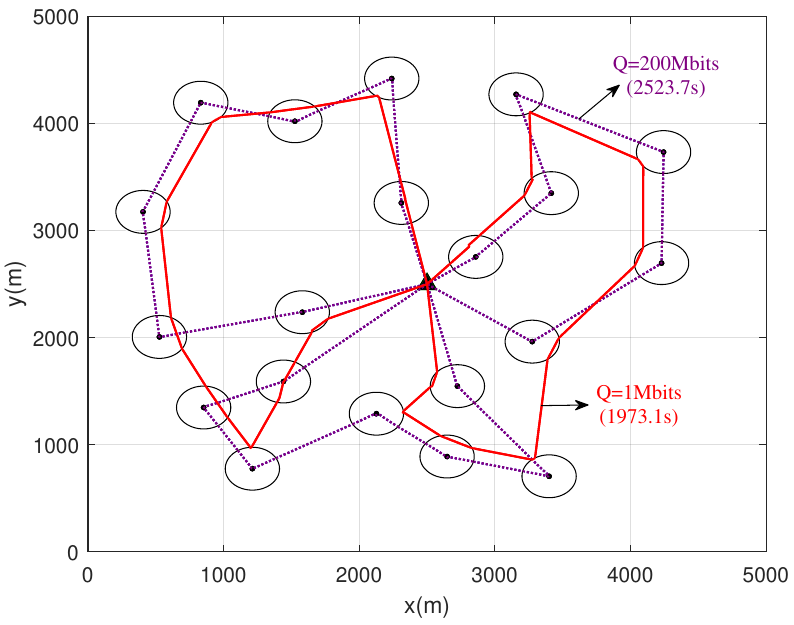}}
    \vspace{-3.5mm}
    \caption{UAV trajectory versus the SN communication throughput requirement ($E_{UAV}$=100KJ, $d_{th}=200m$).}
    \label{trajectory vs Q}
\end{figure}

\begin{figure}[t]
    \centering{\includegraphics[width=2.3in]{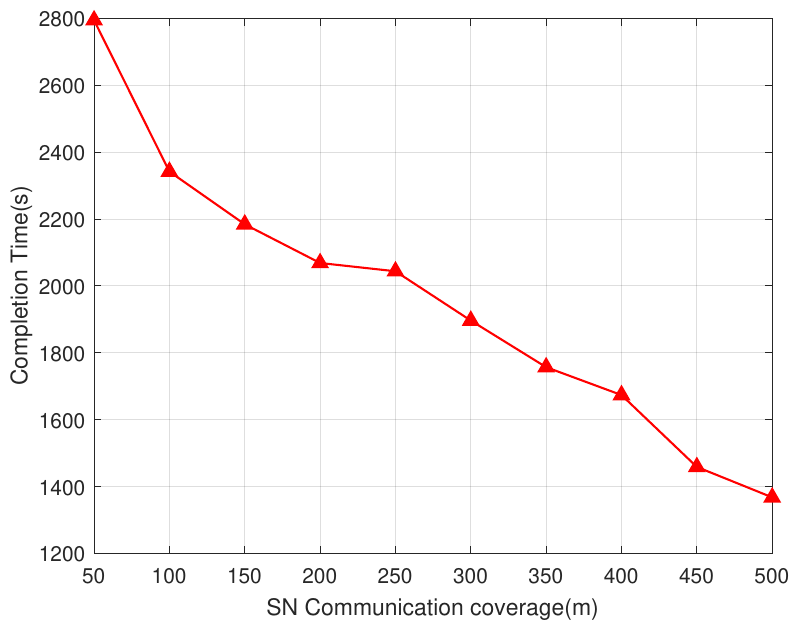}}
    \vspace{-3.5mm}
    \caption{Completion time versus the SN coverage ($Q$=100Mbits, $E_{UAV}$=100KJ).}
    \label{TvsR}
\end{figure}

\begin{figure}[t]
    \centering{\includegraphics[width=2.3in]{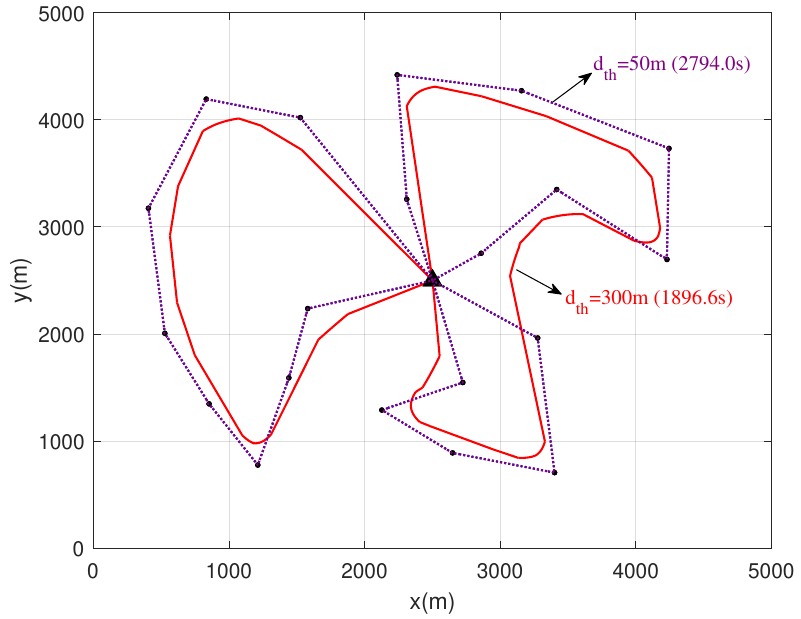}}
    \vspace{-3.5mm}
    \caption{UAV trajectory versus the SN coverage ($Q$=100Mbits, $E_{UAV}$=100KJ).}
    \label{trajectory vs R}
\end{figure}

{\color{black}
As shown in Fig. \ref{TvsSN}, the completion time gradually increases as the number of SNs increases from 20 to 45. 
The completion times of the proposed algorithm at different numbers of SNs are on average about 45.8\% and 38.1\% lower than that of the two benchmarks, respectively.
In addition, it can be seen that the completion time achieved by the proposed PTO algorithm is very close to the lower bound of (P1), the average gap is about 3.9\%. In particular, when the number of SNs is 30, the difference between the completion time of the proposed PTO algorithm and the lower bound is 0.7\%. }

It is observed from Fig. \ref{TvsE} that the completion time achieved by the proposed algorithm under the two benchmarks decreases as the UAV's maximum energy increases since the UAV can collect more SNs in one flight.
Under different UAV maximum energy, the completion time of the PTO algorithm is on average 38.9\% and 34.0\% lower than the benchmarks, respectively. 
As the UAV's maximum energy continues to increase, the SN clusters and collection order are fixed, resulting in convergence of completion time.

Fig. \ref{TvsQ} shows the completion time for different communication throughput requirements. 
It can be seen that with the increase in communication throughput demand, the completion time also increases. 
The main reason is the increase in time for the UAV to collect data from SNs and the recharging time caused by higher energy consumption.
Compared with the greedy algorithm and the optimized Hmode algorithm, the completion time of the proposed PTO algorithm is reduced by 39.1\% and 32.0\% on average, respectively.

{\color{black}To exhibit the computational cost of the proposed algorithm, we compute the execution time and compare it with the benchmarks.
Table \ref{time} presents the execution time for different algorithms, measured on an Intel Core i5 processor with a CPU clock speed of 2.60 GHz.
As the number of SNs increases, the execution time of the algorithms increase gradually.
It can be found that the running time of the proposed PTO algorithm is slightly higher than the benchmarks. 
The main reason is that the PTO algorithm optimizes the UAV trajectory in the SNs' communication coverage while the benchmarks are both hovering above the SNs to collect data.
Although the execution time is slightly higher, the PTO algorithm can obtain  more optimal flight trajectory and data collection completion time within a few iterations.
Furthermore, since the algorithm is executed offline before the UAV flies rather than during flight, it is acceptable for obtaining more optimal trajectory with slightly increased execution time. }

It can be seen from Fig. \ref{trajectory vs E} that the trajectory of the UAV differs under the different energy limitations and the same communication requirements $Q$=150Mbits. 
The energy of the UAV affects the number of times the UAV returns to the charging platform for charging. 
When the UAV's energy is 50KJ, the UAV needs to return 9 times to complete the data collection of all SNs, while when the UAV's energy increases to 100KJ, as shown by the red solid line trajectory in the figure, the UAV only needs to return to the charging platform 3 times, greatly reducing the total task completion time.

Fig. \ref{trajectory vs Q} shows the UAV trajectory under the different communication requirements and the same energy limitations $E_{UAV}$=100KJ. 
It can be seen that the amount of data affects the UAV flight trajectory within the SN's communication coverage. 
When the data amount is large ($Q$=200Mbits), the UAV tends to fly directly above the SN, because this can improve the communication rate and reduce the data collection time. 
However, when the amount of data is small ($Q$=1Mbits), the UAV only needs to fly over the edge of the SN communication coverage to meet the communication requirements, thus reducing the total flight distance and time.

{\color{black}The remaining results are obtained with $Q$=100Mbits and $E_{UAV}$=100KJ.}
Fig. \ref{TvsR} and Fig. \ref{trajectory vs R} show the total completion time and UAV trajectory during the data collection mission under different SN communication coverages.
It can be seen that as the SN communication coverage increases, the total task completion time decreases. 
The main reason is that when the communication coverage is larger, the UAV can start the data collection earlier, and it does not need to fly closer to the SN to complete the data collection task, which reduces the total flight distance and therefore the flight time. 
In addition, the flight energy consumption is also reduced, allowing the UAV to collect data from more SNs at once, then reducing charging time, and ultimately reducing the completion time.

\begin{figure}[t]
    \centering{\includegraphics[width=2.3in]{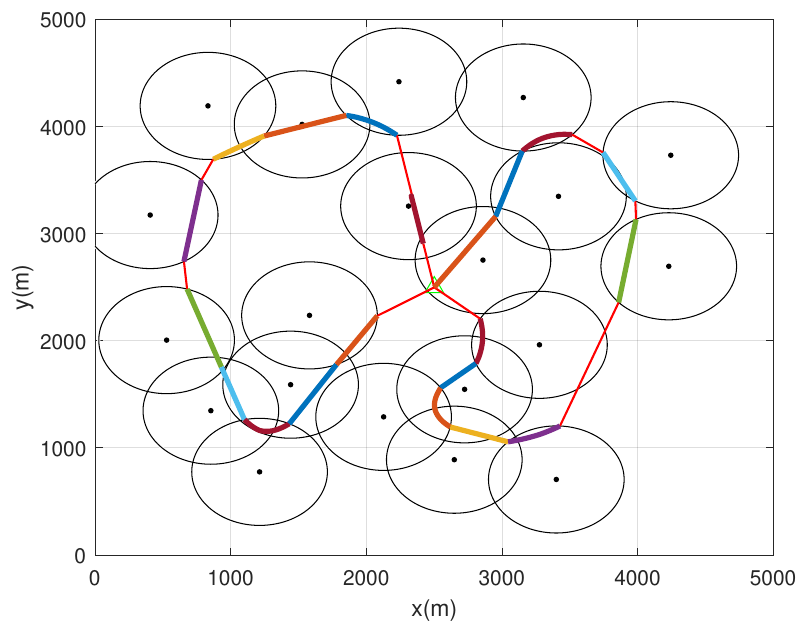}}
    \vspace{-3.5mm}
    \caption{UAV trajectory under overlapped SNs' communication coverage ($Q$=100Mbits, $E_{UAV}$=100KJ, $d_{th}=500m$).}
    \label{R500}
\end{figure}

Furthermore, as the SNs' communication coverage increases, the overlapped communication areas will appear, as shown in Fig. \ref{R500}, the red line represents the flight path between the SNs, and the color bold curves represent the data collection trajectories for different SNs. It can be seen that in the overlapped areas, the UAV can select different SNs for data collection flexibly through the proposed algorithm, thus obtaining a shorter path and completion time.

\begin{figure}[t]
    \centering{\includegraphics[width=2.3in]{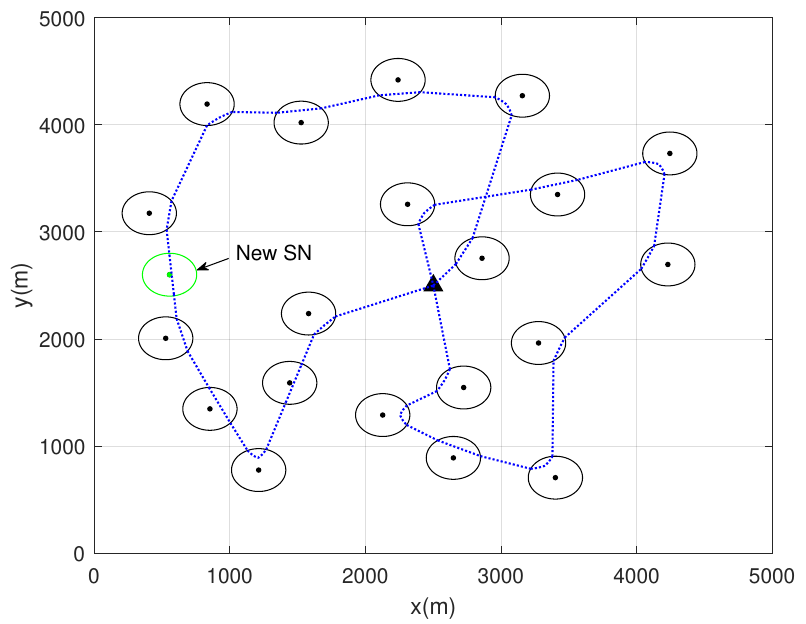}}
    \vspace{-3.5mm}
    \caption{SN addition ($Q$=100Mbits, $E_{UAV}$=100KJ, $d_{th}=200m$).}
    \label{NewSN}
\end{figure}

\begin{figure}[t]
    \centering{\includegraphics[width=2.3in]{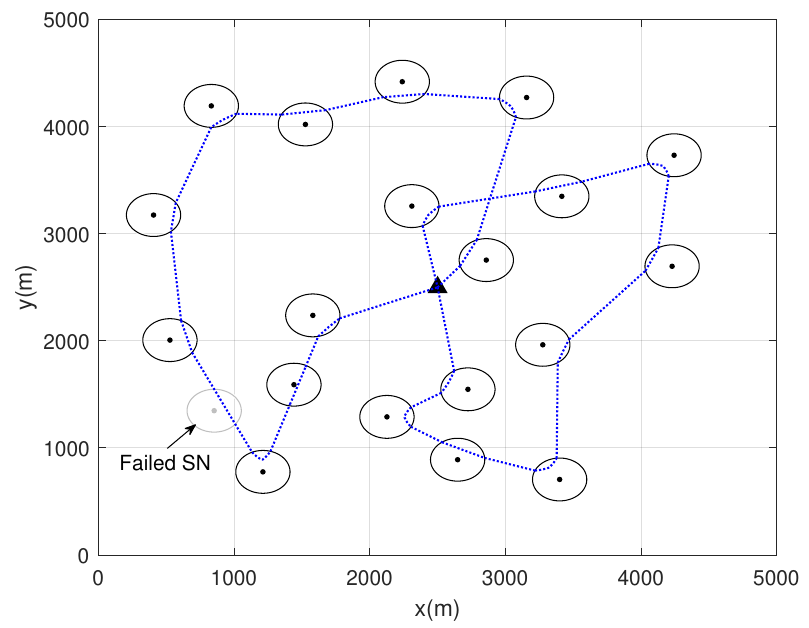}}
    \vspace{-3.5mm}
    \caption{SN failure ($Q$=100Mbits, $E_{UAV}$=100KJ, $d_{th}=200m$).}
    \label{Failed SN}
\end{figure}

For the dynamic scenarios with SNs change, when the new SN's communication coverage intersects the optimized trajectory, as shown in Fig. \ref{NewSN}, if the condition $\Tilde{Q}\geq Q$ is satisfied, then the optimal trajectory of the new problem is the same as that of the original problem (c.f. Theorem 3 and Fig. \ref{different trajectories}(c)). In Fig. \ref{Failed SN}, a similar result is also shown when SN fails at the straight line of the original optimized trajectory.

\section{Conclusion}
This paper investigated the UAV-assisted persistent data collection in large-scale sensor networks.
We proposed a rechargeable UAV-assisted periodic data collection scheme, and formulated an optimization problem aimed at minimizing the periodic completion time by optimizing the UAV data collection trajectory, the SNs subset selection, and the data collection sequence.
The formulated problem was decomposed into two sub-problems and addressed by leveraging SCA, bisection, and heuristic method. Then, we proposed a periodic trajectory optimization algorithm to iteratively solve the two sub-problems to minimize the completion time.
Furthermore, to deal with the dynamics of SNs, we proposed a low-complexity trajectory adjustment strategy that can significantly mitigates the computation cost of recalculation.
The simulation results showed the superiority and robustness of the proposed scheme and the completion time is on average 39\% and 33\% lower than the two benchmarks, respectively.
In the future, we expect to investigate the multi-UAV trajectory optimization and charging strategy for real-time data collection in large-scale mobile sensor scenarios, which can enhance the universality and applicability in diverse application scenarios.

\appendices

\begin{figure}[t]
    \centering{\includegraphics[width=3.1in]{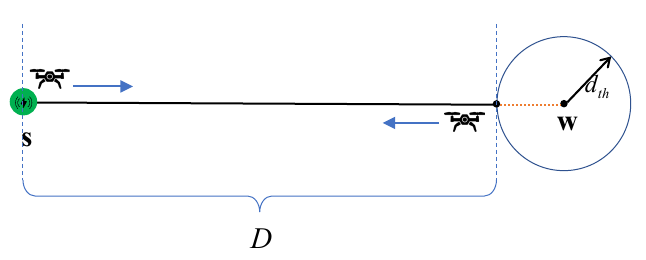}}
    \vspace{-3.5mm}
    \caption{The feasibility condition of (P1).}
    \label{feasibility}
\end{figure}

\section{Proof of Proposition 1}
To enable the UAV to complete the task of data collection, the distance between the SN and the charging platform should be limited. Otherwise, when the distance is too large, the UAV may not have enough energy to return after collecting data or even fail to collect data. Therefore, as shown in Fig. \ref{feasibility}, the UAV should meet the following requirements when collecting data,
\vspace{-1.5mm}
\begin{equation}
    E_{ad} + 2P(V_f)\frac{D}{V_f} + E_{com} \leq E_{UAV}.
    \vspace{-1.5mm}
\end{equation}
Then, we can obtain
\vspace{-1.5mm}
\begin{equation}
    D \leq (E_{UAV}-E_{ad}-E_{com})\frac{V_f}{2P(V_f)}.
    \vspace{-1.5mm}
\end{equation}
Since $E_{UAV}$, $E_{ad}$, and $V_f$ are constants, the maximum value of $D$ should be obtained when $E_{com}$ takes the minimum value. The minimum value of $E_{com}$ can be obtained by solving the following problem,
\vspace{-1.5mm}
\begin{align}
    & \min_{\{\mathbf{q}[m]\}, \{t[m]\}} E_{com}\\
    \mathrm{s.t.\ }
    & \text{(11a)-(11f)}. \notag
    \vspace{-1.5mm}
\end{align}

The above problem can be solved in the same way as Section \uppercase\expandafter{\romannumeral3}-A. Denote by $E_{com}^{*}$ the optimal solution to the problem. Therefore, (P1) is feasible if and only if $ D \leq (E_{UAV}-E_{ad}-E_{com}^{*}(d_{th},Q_k))\frac{V_{f}}{2P(V_{f})}, \forall k $, which completes the proof.

In addition, since $d_{th}$ is usually much smaller than $D$ in a real scenario, then $d_{th}$ can approximately be 0. In this case, the collection energy can be expressed as $\frac{Q_k P_h}{B \log_2\left(1+\frac{P_t \beta_0}{\sigma^2 H^2}\right)}$, then the feasibility condition of (P1) can be written as $D \leq (E_{UAV}-E_{ad}-\frac{Q_k P_h}{B \log_2\left(1+\frac{P_t \beta_0}{\sigma^2 H^2}\right)})\frac{V_{f}}{2P(V_{f})}, \forall k $, where $P_h = P_0+P_i$, is the UAV hovering power.

\section{Proof of Proposition 2}
When the UAV has sufficient energy, the UAV can complete the data collection task of all SNs in one flight. In this case, the minimum energy consumption of the UAV can be used to calculate the lower bound of the number of charging times of the UAV returning to the charging platform in the actual situation.
Therefore, a UAV energy minimization problem for data collection in one flight is formulated as follows:
\vspace{-1.5mm}
\begin{align}
    &\min_{\{\pi(k)\}, \{\mathbf{q}_{k}[m]\}, \{t_{k}[m]\}} E^{1-flight}_{tot} \\
    \mathrm{s.t.\ }
    & \text{(4), (9a)-(9e), (9g)}. \notag
    \vspace{-1.5mm}
\end{align}

Similar to (P1), the objective function is replaced by the total energy consumption, and $N = 1, l = k, G_n = \mathcal{K}$. Although this problem is non-convex, it can also be solved by leveraging the TSP solution for ${\pi(k)}$ and SCA method for$\{\{\mathbf{q}_{k}[m]\}, \{t_{k}[m]\}\}$. Denoted by $(E_{tot}^{1-flight})^*$ the optimal solution to this problem, then the lower bound of the UAV charging times is
\vspace{-1.5mm}
\begin{equation}
    N^{lb} = \left \lceil \frac{(E_{tot}^{1-flight})^*}{E_{UAV}} \right \rceil,
    \vspace{-1.5mm}
\end{equation}
which completes the proof.

\bibliographystyle{IEEEtran}
\bibliography{IEEEabrv,reference}

\begin{thebibliography}{10}
\providecommand{\url}[1]{#1}
\csname url@samestyle\endcsname
\providecommand{\newblock}{\relax}
\providecommand{\bibinfo}[2]{#2}
\providecommand{\BIBentrySTDinterwordspacing}{\spaceskip=0pt\relax}
\providecommand{\BIBentryALTinterwordstretchfactor}{4}
\providecommand{\BIBentryALTinterwordspacing}{\spaceskip=\fontdimen2\font plus
\BIBentryALTinterwordstretchfactor\fontdimen3\font minus
  \fontdimen4\font\relax}
\providecommand{\BIBforeignlanguage}[2]{{%
\expandafter\ifx\csname l@#1\endcsname\relax
\typeout{** WARNING: IEEEtran.bst: No hyphenation pattern has been}%
\typeout{** loaded for the language `#1'. Using the pattern for}%
\typeout{** the default language instead.}%
\else
\language=\csname l@#1\endcsname
\fi
#2}}
\providecommand{\BIBdecl}{\relax}
\BIBdecl

\bibitem{wang2024rechargeable}
R.~Wang, D.~Li, and K.~Meng, ``{R}echargeable {UAV} trajectory optimization for
  real-time persistent data collection of large-scale sensor networks,'' in
  \emph{2024 IEEE International Conference on Communications Workshops (ICC
  Workshops)}, Jun. 2024, pp. 1481--1486.

\bibitem{Ferrag2023edge}
M.~A. Ferrag, O.~Friha, B.~Kantarci, N.~Tihanyi, L.~Cordeiro, M.~Debbah,
  D.~Hamouda, M.~Al-Hawawreh, and K.-K.~R. Choo, ``{E}dge learning for
  {6G}-enabled internet of things: {A} comprehensive survey of vulnerabilities,
  datasets, and defenses,'' \emph{{IEEE} Commun. Surveys Tuts.}, vol.~25,
  no.~4, pp. 2654--2713, Sep. 2023.

\bibitem{Shen2023fair}
Q.~Shen, J.~Peng, W.~Xu, Y.~Sun, W.~Liang, L.~Chen, Q.~Zhao, and X.~Jia,
  ``{F}air communications in {UAV} networks for rescue applications,''
  \emph{{IEEE} Internet Things J.}, vol.~10, no.~23, pp. 21\,013--21\,025, Dec.
  2023.

\bibitem{meng2023throughput}
K.~Meng, Q.~Wu, S.~Ma, W.~Chen, K.~Wang, and J.~Li, ``Throughput maximization
  for {UAV}-enabled integrated periodic sensing and communication,''
  \emph{{IEEE} Trans. Wireless Commun.}, vol.~22, no.~1, pp. 671--687, Jan.
  2023.

\bibitem{meng2023multi}
K.~Meng, X.~He, Q.~Wu, and D.~Li, ``Multi-{UAV} collaborative sensing and
  communication: {J}oint task allocation and power optimization,'' \emph{{IEEE}
  Trans. Wireless Commun.}, vol.~22, no.~6, pp. 4232--4246, Jun. 2023.

\bibitem{Zhu2022Multi-UAV}
L.~Zhu, J.~Zhang, Z.~Xiao, X.-G. Xia, and R.~Zhang, ``Multi-{UAV} aided
  millimeter-wave networks: Positioning, clustering, and beamforming,''
  \emph{{IEEE} Trans. Wireless Commun.}, vol.~21, no.~7, pp. 4637--4653, Jul.
  2022.

\bibitem{Zhu2020Millimeter}
L.~Zhu, J.~Zhang, Z.~Xiao, X.~Cao, X.-G. Xia, and R.~Schober, ``Millimeter-wave
  full-duplex {UAV} relay: Joint positioning, beamforming, and power control,''
  \emph{{IEEE} J. Sel. Areas Commun.}, vol.~38, no.~9, pp. 2057--2073, Sep.
  2020.

\bibitem{Zhu20193-D}
L.~Zhu, J.~Zhang, Z.~Xiao, X.~Cao, D.~O. Wu, and X.-G. Xia, ``{3-D} beamforming
  for flexible coverage in millimeter-wave {UAV} communications,'' \emph{IEEE
  Wireless Commun. Lett.}, vol.~8, no.~3, pp. 837--840, Jun. 2019.

\bibitem{Lin2019kalman}
Z.~Lin, H.~H.~T. Liu, and M.~Wotton, ``{K}alman filter-based large-scale
  wildfire monitoring with a system of {UAVs},'' \emph{{IEEE} Trans. Ind.
  Electron.}, vol.~66, no.~1, pp. 606--615, Jan. 2019.

\bibitem{Sun2019A}
Y.~Sun, Z.~Ding, and X.~Dai, ``{A} user-centric cooperative scheme for
  {UAV}-assisted wireless networks in malfunction areas,'' \emph{{IEEE} Trans.
  Commun.}, vol.~67, no.~12, pp. 8786--8800, Dec. 2019.

\bibitem{Qin2021Task}
Z.~Qin, H.~Wang, Z.~Wei, Y.~Qu, F.~Xiong, H.~Dai, and T.~Wu, ``{T}ask selection
  and scheduling in {UAV}-enabled {MEC} for reconnaissance with time-varying
  priorities,'' \emph{{IEEE} Internet Things J.}, vol.~8, no.~24, pp.
  17\,290--17\,307, Dec. 2021.

\bibitem{mozaffari2019tutorial}
M.~Mozaffari, W.~Saad, M.~Bennis, Y.-H. Nam, and M.~Debbah, ``A tutorial on
  {UAV}s for wireless networks: {A}pplications, challenges, and open
  problems,'' \emph{{IEEE} Commun. Surveys Tuts.}, vol.~21, no.~3, pp.
  2334--2360, Mar. 2019.

\bibitem{Baek2020energy}
J.~Baek, S.~I. Han, and Y.~Han, ``{E}nergy-efficient {UAV} routing for wireless
  sensor networks,'' \emph{{IEEE} Trans. Veh. Technol.}, vol.~69, no.~2, pp.
  1741--1750, Feb. 2020.

\bibitem{Jia2019age}
Z.~Jia, X.~Qin, Z.~Wang, and B.~Liu, ``{A}ge-based path planning and data
  acquisition in {UAV}-assisted {IoT} networks,'' in \emph{2019 IEEE
  International Conference on Communications Workshops (ICC Workshops)}, 2019,
  pp. 1--6.

\bibitem{liu2022uav}
K.~Liu and J.~Zheng, ``{UAV} trajectory optimization for time-constrained data
  collection in {UAV}-enabled environmental monitoring systems,'' \emph{{IEEE}
  Internet Things J.}, vol.~9, no.~23, pp. 24\,300--24\,314, Dec. 2022.

\bibitem{boukoberine2019critical}
M.~N. Boukoberine, Z.~Zhou, and M.~Benbouzid, ``A critical review on unmanned
  aerial vehicles power supply and energy management: {S}olutions, strategies,
  and prospects,'' \emph{Appl. Energy}, vol. 255, p. 113823, Dec. 2019.

\bibitem{li2018uav}
B.~Li, Z.~Fei, and Y.~Zhang, ``{UAV} communications for 5g and beyond: {R}ecent
  advances and future trends,'' \emph{IEEE Internet of Things Journal}, vol.~6,
  no.~2, pp. 2241--2263, Apr. 2019.

\bibitem{Nguyen2018scheduling}
A.~H. Nguyen, Y.~Tanigawa, and H.~Tode, ``{S}cheduling method for solving
  successive contentions of heterogeneous periodic flows based on mathematical
  formulation in multi-hop {WSNs},'' \emph{IEEE Sensors J.}, vol.~18, no.~21,
  pp. 9021--9033, Nov. 2018.

\bibitem{xu2021event}
S.~Xu, Y.~Liu, W.~Hu, Y.~Wu, S.~Liu, Y.~Wang, and C.~Liu, ``{E}vent-sensitive
  network: {A} construction algorithm of agricultural sensor network driven by
  environmental change,'' \emph{Mathematical Problems in Engineering}, vol.
  Jan. 2021, pp. 1--14, 2021.

\bibitem{javaid2019machine}
A.~Javaid, N.~Javaid, Z.~Wadud, T.~Saba, O.~E. Sheta, M.~Q. Saleem, and M.~E.
  Alzahrani, ``Machine learning algorithms and fault detection for improved
  belief function based decision fusion in wireless sensor networks,''
  \emph{Sensors}, vol.~19, no.~6, p. 1334, Mar. 2019.

\bibitem{zhang2019wireless}
Z.~Zhang, H.~Pang, A.~Georgiadis, and C.~Cecati, ``Wireless power transfer—an
  overview,'' \emph{IEEE Trans. Ind. Electron.}, vol.~66, no.~2, pp.
  1044--1058, Feb. 2019.

\bibitem{ke2017design}
D.~Ke, C.~Liu, C.~Jiang, and F.~Zhao, ``{D}esign of an effective wireless air
  charging system for electric unmanned aerial vehicles,'' in \emph{2017 43rd
  Annual Conference of the IEEE Industrial Electronics Society (IECON)}, 2017,
  pp. 6949--6954.

\bibitem{zhao2020efficiency}
M.-M. Zhao, Q.~Shi, and M.-J. Zhao, ``{E}fficiency maximization for
  {UAV}-enabled mobile relaying systems with laser charging,'' \emph{{IEEE}
  Trans. Wireless Commun.}, vol.~19, no.~5, pp. 3257--3272, May. 2020.

\bibitem{li2021minimizing}
M.~Li, L.~Liu, Y.~Gu, Y.~Ding, and L.~Wang, ``{M}inimizing energy consumption
  in wireless rechargeable {UAV} networks,'' \emph{{IEEE} Internet Things J.},
  vol.~9, no.~5, pp. 3522--3532, Mar. 2021.

\bibitem{MOHAMMADNIA2021107283}
A.~Mohammadnia, B.~{M. Ziapour}, H.~Ghaebi, and M.~H. Khooban, ``{F}easibility
  assessment of next-generation drones powering by laser-based wireless power
  transfer,'' \emph{Opt. Laser. Technol.}, vol. 143, p. 107283, Jun. 2021.

\bibitem{li2019rechargeable}
X.~Li, H.~Yao, J.~Wang, S.~Wu, C.~Jiang, and Y.~Qian, ``{R}echargeable
  multi-{UAV} aided seamless coverage for {QoS}-guaranteed {IoT} networks,''
  \emph{IEEE Internet of Things Journal}, vol.~6, no.~6, pp. 10\,902--10\,914,
  Sep. 2019.

\bibitem{chen2022trajectory}
W.~Chen, D.-K. Chang, and Y.-J. Chen, ``{T}rajectory control in
  self-sustainable {UAV}-aided mmwave networks: {A} constrained multi-agent
  reinforcement learning approach,'' in \emph{2022 IEEE International
  Conference on Communications Workshops (ICC Workshops)}.\hskip 1em plus 0.5em
  minus 0.4em\relax IEEE, 2022, pp. 1017--1022.

\bibitem{zhu2021efficient}
Y.~Zhu and S.~Wang, ``{E}fficient aerial data collection with cooperative
  trajectory planning for large-scale wireless sensor networks,'' \emph{{IEEE}
  Trans. Commun.}, vol.~70, no.~1, pp. 433--444, Jan. 2022.

\bibitem{zhu2022aerial}
K.~Zhu, J.~Yang, Y.~Zhang, J.~Nie, W.~Y.~B. Lim, H.~Zhang, and Z.~Xiong,
  ``{A}erial refueling: Scheduling wireless energy charging for {UAV} enabled
  data collection,'' \emph{IEEE Transactions on Green Communications and
  Networking}, vol.~6, no.~3, pp. 1494--1510, Sep. 2022.

\bibitem{zhang2021energy}
L.~Zhang, A.~Celik, S.~Dang, and B.~Shihada, ``{E}nergy-efficient trajectory
  optimization for {UAV}-assisted {IoT} networks,'' \emph{{IEEE} Trans. Mobile
  Comput.}, vol.~21, no.~12, pp. 4323--4337, Dec. 2022.

\bibitem{virgili2022cost}
M.~Virgili, N.~Babu, M.~Javidsharifi, I.~Valiulahi, C.~Masouros, A.~J. Forsyth,
  T.~Kerekes, and C.~B. Papadias, ``{C}ost-efficient design of an
  energy-neutral {UAV}-based mobile network,'' \emph{{IEEE} Trans. Commun.},
  vol.~70, no.~10, pp. 6890--6901, Oct. 2022.

\bibitem{Muruganathan2021}
S.~D. Muruganathan, X.~Lin, H.-L. Määttänen, J.~Sedin, Z.~Zou, W.~A.
  Hapsari, and S.~Yasukawa, ``An overview of {3GPP} release-15 study on
  enhanced {LTE} support for connected drones,'' \emph{IEEE Commun. Mag.},
  vol.~5, no.~4, pp. 140--146, Dec. 2021.

\bibitem{liao2022energy}
Y.~Liao and V.~Friderikos, ``Energy and age pareto optimal trajectories in
  {UAV}-assisted wireless data collection,'' \emph{{IEEE} Trans. Veh.
  Technol.}, vol.~71, no.~8, pp. 9101--9106, Aug. 2022.

\bibitem{zeng2019energy}
Y.~Zeng, J.~Xu, and R.~Zhang, ``Energy minimization for wireless communication
  with rotary-wing {UAV},'' \emph{{IEEE} Trans. Wireless Commun.}, vol.~18,
  no.~4, pp. 2329--2345, Apr. 2019.

\bibitem{BRAEKERS2016300}
K.~Braekers, K.~Ramaekers, and I.~{Van Nieuwenhuyse}, ``The vehicle routing
  problem: State of the art classification and review,'' \emph{Comput. Ind.
  Eng.}, vol.~99, pp. 300--313, Sep. 2016.

\bibitem{6031159}
I.~Kara, ``Arc based integer programming formulations for the distance
  constrained vehicle routing problem,'' in \emph{3rd IEEE International
  Symposium on Logistics and Industrial Informatics}, Aug. 2011, pp. 33--38.

\bibitem{laporte1987branch}
G.~Laporte, Y.~Nobert, and S.~Taillefer, ``{A} branch-and-bound algorithm for
  the asymmetrical distance-constrained vehicle routing problem,''
  \emph{Mathematical Modelling}, vol.~9, no.~12, pp. 857--868, 1987.

\bibitem{marks1978general}
B.~R. Marks and G.~P. Wright, ``{A} general inner approximation algorithm for
  nonconvex mathematical programs,'' \emph{Oper. Res.}, vol.~26, no.~4, pp.
  681--683, Aug. 1978.

\bibitem{almoustafa2013new}
S.~Almoustafa, S.~Hanafi, and N.~Mladenovi{\'c}, ``{N}ew exact method for large
  asymmetric distance-constrained vehicle routing problem,'' \emph{Eur. J.
  Oper. Res.}, vol. 226, no.~3, pp. 386--394, 2013.

\bibitem{meng2021space}
K.~Meng, D.~Li, X.~He, and M.~Liu, ``{S}pace pruning based time minimization in
  delay constrained multi-task {UAV}-based sensing,'' \emph{{IEEE} Trans. Veh.
  Technol.}, vol.~70, no.~3, pp. 2836--2849, Mar. 2021.

\bibitem{jonker1983transforming}
R.~Jonker and T.~Volgenant, ``{T}ransforming asymmetric into symmetric
  traveling salesman problems,'' \emph{Oper. Res. Lett.}, vol.~2, no.~4, pp.
  161--163, 1983.

\bibitem{LKH}
K.~Helsgaun, ``An effective implementation of the lin–kernighan traveling
  salesman heuristic,'' \emph{Eur. J. Oper. Res.}, vol. 126, no.~1, pp.
  106--130, Oct. 2000.

\bibitem{li2023completion}
M.~Li, X.~Liu, and H.~Wang, ``Completion time minimization considering {GN}s’
  energy for {UAV}-assisted data collection,'' \emph{{IEEE} Commun. Lett.},
  vol.~12, no.~12, pp. 2128--2132, Dec. 2023.

\end{thebibliography}

\end{document}